\documentclass[journal]{IEEEtran}

\usepackage{cite}
\usepackage{amsmath,amssymb,amsfonts}
\usepackage{algorithm,algorithmic}
\usepackage{graphicx}
\usepackage{textcomp}
\usepackage{color,xcolor}
\usepackage{bm}
\usepackage{subfigure}
\usepackage{multirow}
\usepackage[normalem]{ulem}
\usepackage{tabularx}
\usepackage{xpatch}
\usepackage{soul}
\usepackage{ulem}
\usepackage{verbatim}
\usepackage{geometry}

\ifCLASSINFOpdf
\else
\fi

\begin{document}


\title{Angular-Distance Based Channel Estimation for Holographic MIMO}

\author{
	Yuanbin~Chen,~Ying~Wang,~\IEEEmembership{Member,~IEEE,}~Zhaocheng~Wang,~\IEEEmembership{Fellow,~IEEE,} \\ and~Zhu~Han,\IEEEmembership{~Fellow,~IEEE}

\thanks{
	This work was supported in part by the National Key R\&D Program of China under grant 2022YFA1003901,
	in part by the Beijing Natural
	Science Foundation under Grant 4222011, and in part by NSF CNS-2107216, CNS-2128368, CMMI-2222810, ECCS-2302469, US Department of Transportation, Toyota and Amazon. \textit{(Corresponding authors: Ying Wang; Zhaocheng Wang.)}
	
	Yuanbin Chen and Ying Wang are with the State Key Laboratory of Networking and Switching Technology, Beijing University of Posts and Telecommunications, Beijing 100876, China (e-mail: chen\_yuanbin@163.com; wangying@bupt.edu.cn).
	
	Zhaocheng Wang is with the Beijing National Research Center for Information Science and Technology, Department of Electronic Engineering, Tsinghua University, Beijing 100084,  China (e-mail: zcwang@tsinghua.edu.cn).
	
	Zhu Han is with the Department of Electrical and Computer Engineering in the University of Houston, Houston, TX~77004, USA, and also with the Department of Computer Science and Engineering, Kyung Hee University, Seoul, South Korea, 446-701 (e-mail: hanzhu22@gmail.com).
}

}

%



\maketitle

\begin{abstract}
Leveraging the concept of the electromagnetic signal and information theory, holographic multiple-input multiple-output (MIMO) technology opens the door to an intelligent and endogenously holography-capable wireless propagation environment, with their unparalleled capabilities for achieving high spectral and energy efficiency. Less examined are the important issues such as the acquisition of accurate channel information by accounting for holographic MIMO’s peculiarities. To fill this knowledge gap, this paper investigates the channel estimation for holographic MIMO systems by unmasking their distinctions from the conventional one. Specifically, we elucidate that the channel estimation, subject to holographic MIMO's electromagnetically large antenna arrays, has to discriminate not only the angles of a user/scatterer but also its distance information, namely the three-dimensional (3D) azimuth and elevation angles plus the distance (AED) parameters. As the angular-domain representation fails to characterize the sparsity inherent in holographic MIMO channels, the tightly coupled 3D AED parameters are firstly decomposed for independently constructing their own covariance matrices. Then, the recovery of each individual parameter can be structured as a compressive sensing (CS) problem by harnessing the covariance matrix constructed. This pair of techniques contribute to a parametric decomposition and compressed deconstruction (DeRe) framework, along with a formulation of the maximum likelihood estimation for each parameter.
Then, an efficient algorithm, namely DeRe-based variational Bayesian inference and message passing (DeRe-VM), is proposed for the sharp detection of the 3D AED parameters and the robust recovery of sparse channels.
Finally, the proposed channel estimation regime is confirmed to be of great robustness in accommodating different channel conditions, regardless of the near-field and far-field contexts of a holographic MIMO system, 
as well as an improved performance in comparison to the state-of-the-art benchmarks.
\end{abstract}

\begin{IEEEkeywords}
Holographic MIMO, electromagnetic signal, channel estimation, near-field communications.
\end{IEEEkeywords}

%
\IEEEpeerreviewmaketitle

\section{Introduction}

The paradigm shift from the fifth-generation (5G) to the sixth-generation (6G) communications catalyzes pillar functionalities shared among plethora of newly encountered particularities, with critical performance metrics to be placated including but not subject to ultra-wideband, ultra-massive access, ultra-reliability, and low latency \cite{6G-CST,NC,Michail}. As a ``holly grail” of wireless technologies, massive multi-input and multi-output (mMIMO) has non-trivial capabilities of achieving unprecedented spectral efficiency, the critical ingredients of which have been commercially rolled out and approved for inclusion in 5G next radio (NR) standards \cite{3GPP-R17}. While looking back from MIMO, mMIMO, to the more recent extremely large-scale (XL) MIMO, one may notice that all three adhere to the same philosophy, namely to adapt to an uncontrollable wireless environment. Their beamforming functionalities conform to the beam-space model in the specific domain based upon a number of ideal assumptions, e.g., predefined antenna array geometry with explicit calibrations, along with propagation in the absence of near-field scattering and mutual coupling \cite{deLamare,KV,Edward,hanzhu-TCCN,chen-jsac}. This would be fruitless as the array aperture becomes larger and electromagnetically manipulated, shaped in arbitrary geometries and covered with dense antenna elements. Therefore, we are highly anticipating the evolution of the classical signal processing methodologies to the electromagnetic (EM) level using EM signal and information theory, allowing for greater degrees of freedom in the context of 6G~\cite{GK}.

Benefiting from the advancement of meta-materials and meta-surfaces, along with their widespread uses in wireless communications, holographic MIMO has been crafted as an excellent candidate to enable holographic radios. The fundamentals of holographic MIMO have been expounded in \cite{Holo-101,Holo-102,Holo-105}, and herein we will shine the spotlight on the qualitative leaps from the conventional MIMO to its holographic counterpart for the following twofold aspects. Firstly, regarding physical fundamentals, holographic MIMO is considered to have a nearly continuous aperture with antenna element spacing being much less than half a wavelength of incident EM waves. This contrasts with the half-wavelength condition typical in conventional MIMO. Such a distinction leads to transformative changes in hardware structure, notably in terms of the array aperture size (from small to extremely large) and the number/density of antenna elements (from sparse to dense). Intriguingly, the densely packed antenna elements may lead to enhancement in spectral efficiency \cite{Holo_25}. Furthermore, the attendant mutual coupling among antenna elements, albeit deemed detrimental in conventional MIMO designs, fundamentally transforms, and is potential to be leveraged in achieving the super-directivity, a phenomenon that describes the significantly higher array gains obtained in holographic MIMO~\cite{Marzetta,Holo_26}.
Secondly, from the perspective of propagation channels, another qualitative change in holographic MIMO results from its extremely large aperture size. Unlike mMIMO communications that typically employ planar wavefront in far-field cases, holographic MIMO can naturally draw the far-field region to its near-field counterpart (also known as the Fresnel region) as the aperture size increases. Therefore, holographic MIMO near-field communications have to discriminate not only the angle of a scatterer but also its distance information, in contrast to the angle-only-aware counterpart in mMIMO systems.

Given that holographic MIMO just plays a basic physical role, associated physical-layer technologies are required for further implementation of holographic communications. To this end, significant efforts have been devoted to the channel modeling \cite{Holo-2,Holo-3,Holo-4,Holo-13,Holo-14}, channel estimation \cite{Holo-4, Holo-15}, and holographic beampattern design \cite{Holo-9, Holo-10, Holo-5}. In these literatures, the holographic channel models employed might not arrive at a good consensus, which can be roughly classified into two categories: i) EM channel based on Fourier planar wavefront representation \cite{Holo-2,Holo-13,Holo-14}, and ii) parametric physical channel based on multi-path representation \cite{Holo-15,XLM-1}. More precisely, the former goes more deeply into characterizing what kind of channel distributions that could appear in the radiative near-field and what cannot. The latter is more concentrated on the imposed sparsity in the sense that the received signal contains a sum of a few spherical waves, based upon which the compressive sensing (CS) method can be harnessed for facilitating high-efficiency parameter recovery. In spite of this, we believe that this pair of somewhat different channel models for characterizing holographic MIMO may be aligned to the same ingredients but using different methods for analyzing the holographic MIMO channels.

Holographic MIMO’s intriguing features contribute significantly to the achievement of holographic radios, yet just a few fledgling efforts on channel estimation in the true sense of the term are available to date, e.g., \cite{Holo-4,Holo-15}. Even so, we are fortunate to observe, as more attention is drawn to related topics, the emergence of pioneering studies with a focus on near-field channel estimation based on the assumption of spherical wavefront  utilizing various techniques~\cite{Holo-15,XLM-1,XLM_A-3, XLM_12,XLM_A-18}, such as CS-based approaches~\cite{XLM-1,XLM_A-18}, and learning-based means~\cite{XLM_12}. In particular, the potential sparsity intrinsic to the near-field channel has been unveiled in \cite{XLM-1}, where a polar-domain representation is proposed in lieu of the angular-domain counterpart in the far field. Base upon the polar-domain philosophy, the near-field codebook solution is devised in \cite{XLM_A-3} for multiple access. In \cite{XLM_A-18}, a joint dictionary learning and sparse recovery empowered channel estimation is presented for near-field communications. Furthermore, in a distinct approach, a model-based deep learning framework for near-field channel estimation is showcased in \cite{XLM_12}, in which the channel estimation problem is attributed to a CS problem that is solved by deep learning-based techniques.

Despite these impressive initiatives, we evince that the channel estimation for the holographic MIMO particularly for the near-field regime remains still an open issue, as evidenced by the following facts. Firstly, owing to the extremely larger array size and denser element spacing, the conventional array modeling, which treats the antenna elements as sizeless points, makes no sense. A generic model is required for naturally transitioning the discrete antenna aperture to its continuous counterpart, for which EM waves can be manipulated by continuous-to-discrete antenna elements that are sub-wavelength spaced for a holographic array. 
Secondly, while the sparse representations presented in \cite{XLM-1,XLM_12}, and \cite{XLM_A-18} indeed tackle the energy spread effect exposed in the near-field scenario, their applicability is confined to the uniform linear array (ULA) case, in the sense that their extensions to more complicated uniform planar array (UPA) regimes are problematic. This is attributed to the intricate coherence between the two-dimensional (2D) azimuth-elevation angle pair and distance parameters as well as a more stringent restrict isometry property (RIP) condition. In \cite{XLM_A-3}, although the UPA case is considered for the near-field codebook design, a simplified assumption is adopted, in which the UPA case is approximatedly pushed to an asymptotic ULA case when conducting coherence analysis. We thus intend to revisit this pressing issue within the context of holographic UPA configurations in this treatise. 
As a third consideration, given the beneficial sparsity in facilitating holographic channel estimation, it is imperative to account for the overhead induced by channel sparsification and the complexity of algorithm design.
The modeling approaches in~\cite{XLM-1}~and~\cite{XLM_A-18} facilitate the relief of heavy overhead to different degrees,
but suffer from the multiplicative gridding complexity, i.e., the product of individual complexity associated with discretizing the azimuth angle, elevation angle, and the distance of holographic MIMO channels. This significantly undermines the efficiency of pursuing the optimum in the CS procedure, thus in need of more efficient techniques for searching through this three-dimensional (3D) azimuth and elevation angles plus the distance (AED) grid. In a nutshell, a channel estimation strategy for holographic MIMO systems has to be tailored in response to these non-trivial challenges, with the goal of encapsulating the potential sparsities for a robust and accurate estimate performance at reduced complexity and overhead.

Geared towards the challenges outlined, the current work set out with the aim of filling the knowledge gap in the state-of-the-art by following contributions:

\begin{itemize}
\item We investigate the uplink channel estimation in the holographic MIMO system. For an electromagnetically large-scale UPA, a unified array model is exploited in order for EM waves to be manipulated by the antenna elements that are sub-wavelength spaced across the holographic array of interest. Furthermore, the employed model portrays the variations of signal amplitude, phase, and projected aperture across array positions, based upon which a parametric channel model is presented. By revealing the invalidity of the employment of conventional planar wavefront in holographic MIMO channels, we evince that the channel estimation has to discriminate not only the angle of a user/scatterer but also its distance information, with to-be-determined 3D AED parameters involved.

\item To facilitate a sharp detection for the 3D AED parameters, we conceive a parametric \textbf{de}composition and compressed \textbf{re}construction (DeRe) framework for holographic MIMO systems. To elaborate, the tightly coupled 3D AED parameters are decomposed by constructing the one-dimensional (1D) covariance matrix associated with the azimuth angle, the elevation angle, and the distance, for facilitating their independent estimates. Furthermore, an off-grid basis is utilized to mine potential sparsities inherent to the covariance matrix constructed, and then the 3D AED parameters can be reconstructed as their respective 1D compressive sensing (CS) problems. To this end, it is sufficient to estimate the offset vectors associated with each dimension. Additionally, we evince that the proposed DeRe paradigm may be a curse when concerning the attendant angular index misinterpretation induced by the improper stitching of the obtained three 1D parameters, but a blessing when embraced with regard to significantly reduced gridding complexity, followed by the angular index correction as a remedy.

\item Due to the ill-conditioned sensing matrix (containing offset vectors) and the sparse vectors with imperfect priors, the formulated 1D CS problem may differ somewhat from the standard one, demonstrating a high level of intractability. To tackle this issue, a layered probability model for unknown parameters is firstly established to capture the specific structure of sparse vectors, with the capability of addressing the uncertainty of imperfect prior information. Based on this, the maximum likelihood estimation (MLE) of the offset vectors can be acquired, along with the minimum mean square error (MMSE) estimation of the sparse vectors. Then, we propose an efficient algorithm, namely DeRe-based variational Bayesian inference and message passing (DeRe-VM) in pursuit of a stable solution in a turbo-like way. The proposed DeRe-VM algorithm is featured by a computational complexity that scales proportionally with the number of antenna elements raised to the power of 1.5. This is markedly more efficient when contrasted with state-of-the-art schemes, where the complexity exhibits a quadratic relationship.

\item Simulation results underscore the role of DeRe framework as a critical instrument in mitigating false alarms (misinterpreting the genuine power of the non-significant paths as their significant counterparts) in the process of parameter detection. The proposed DeRe-VM algorithm exhibits a notable superiority over the state-of-the-art angular-domain solutions, primarily attributable to its use of a more accurate Fresnel approximation, without noticeable increase in gridding complexity. Furthermore, we demonstrate that the proposed DeRe-VM algorithm is featured by significant robustness, not only due to its ability to accommodate various channel conditions -- such as the level of channel sparsity and the near-field or far-field contexts of the holographic MIMO system -- but also in its insensitivity agasint perturbations induced by imperfections in practical implementations, thereby ensuring the accuracy of the parameter recovery.


\end{itemize}

The remainder of this paper is organized as follows. Section~II introduces the system model and the challenges to be resolved. In Section~III, we show the way to achieve the parametric decomposition, and in Section~IV, we reconstruct  the parameters in each dimension into their own CS problems. Then, an efficient DeRe-VM algorithm is proposed in Section~V for attaining stable solutions to the formulated CS problems. Section~VI provides numerical results for performance evaluations, along with some useful insights, and finally the paper is concluded in Section~VII.

\section{System Model}
\subsection{Scenario}

\begin{figure*}[t]
	\centering
	\includegraphics[width=0.9\textwidth]{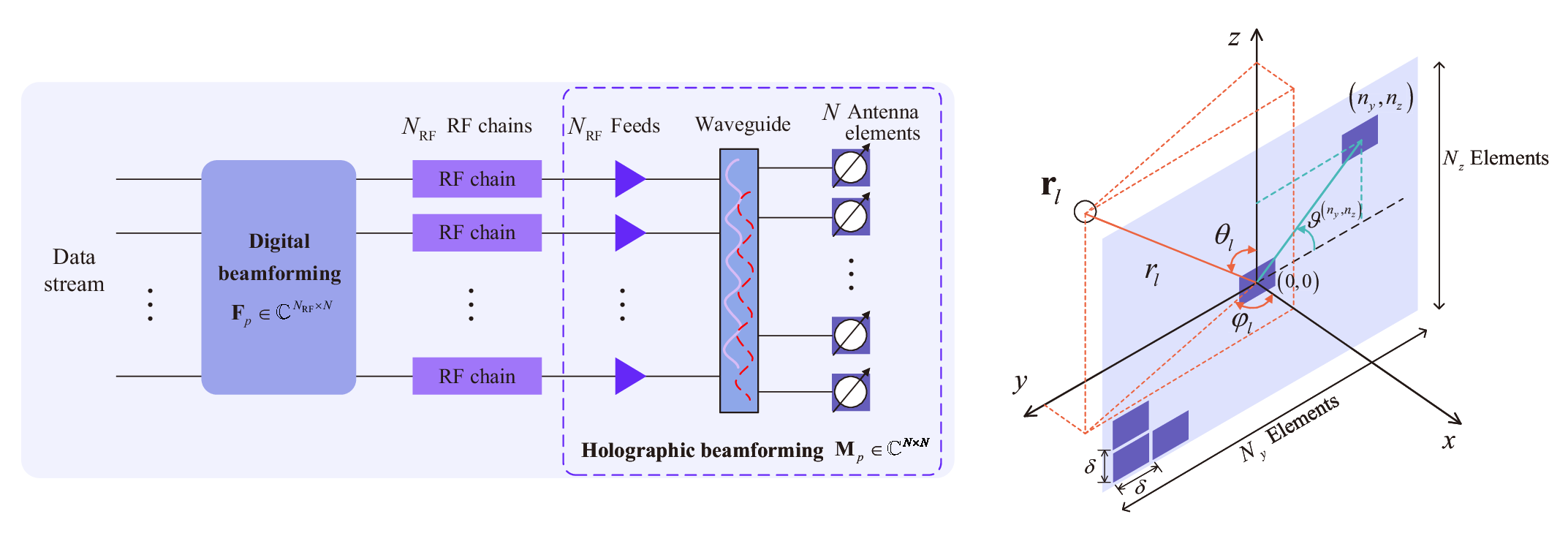}
	\caption{Schematic diagram of holographic MIMO transmission and UPA.} \label{UPA}
\end{figure*}

We consider an uplink holographic MIMO system where multiple single-antenna users are served by a base station (BS) having $N = {N_y} \times {N_z}$ antenna elements in the form of a UPA. ${N_{{\text{RF}}}}\left( {{N_{{\text{RF}}}} \ll N} \right)$ feeds are embedded in the substrate in order to generate reference waves carrying user-intended signals, with each feed connected to a radio frequency (RF) chain for signal processing. 
As illustrated in Fig.~\ref{UPA}, the UPA coincides with the $y$-$z$ plane and centered at the origin, where $N_y$ and $N_z$ denote the number of antenna elements along the $y$- and $z$-axis, respectively. For notational convenience, $N_y$ and $N_z$ are assumed to be odd numbers. The central location of the  $ \left( {{n_y},{n_z}} \right) $th antenna element is $ {{\mathbf{p}}_{{n_y},{n_z}}} = {\left[ {0,{n_y}\delta ,{n_z}\delta } \right]^T} $, in which ${n_y} = \left\{ {0, \pm 1, \pm 2,..., \pm \frac{{{N_y} - 1}}{2}} \right\}$ and ${n_z} = \left\{ {0, \pm 1, \pm 2,..., \pm \frac{{{N_z} - 1}}{2}} \right\}$. Along the $y$- and $z$-axis, the physical measurements of the UPA are approximately $ {N_y}\delta  $  and  $ {N_z}\delta  $, respectively, with $\delta$  representing the inter-element spacing.

In contrast to the conventional array modeling that treats each antenna element as a sizeless point, the size of an antenna element, denoted by  $\sqrt A  \times \sqrt A $, should be explicitly addressed. It is possible to theoretically unify the modeling of continuous holographic surface and discrete antenna arrays by adjusting the element size $A$ and their inter-element spacing $\delta$ \cite{Holo-2, Holo-101}, where  $ \delta  \ge \sqrt A  $. Particularly, let  $ \frac{A}{{{\delta ^2}}} \le 1 $ denote the array occupation ratio that indicates the portion of the entire UPA plate area being occupied by the antenna elements. Regarding the extreme case of  $ \frac{A}{{{\delta ^2}}} = 1 $, the UPA can be treated as a contiguous surface. Therefore, such a modeling is imperative to account for the projected aperture of each antenna element when signals are impinged from various directions, particularly in the presence of extremely large number of antenna elements. Furthermore, we assume in this paper that the effective antenna aperture is  $ A = \frac{{{\lambda ^2}}}{{4\pi }} $ for the isotropic antenna element.

\subsection{Channel Model and Communication Protocol}

We model the channel spanning from the BS to the user based on path parameters of the system \cite{Holo-15,XLM-1}. The assumption of the planar wavefront fails to describe the features of the link distance falling within the range of Rayleigh distance as EM waves do not uniformly radiate across different antenna elements. In this case, the generic spherical-wavefront model is essential to encapsulate phase variations across different antenna elements. In particular, we denote $ {\mathcal{A}_{{n_y},{n_z}}} = \left[ {{n_y}\delta  - \frac{{\sqrt A }}{2},{n_y}\delta  + \frac{{\sqrt A }}{2}} \right] \times \left[ {{n_z}\delta  - \frac{{\sqrt A }}{2},{n_z}\delta  + \frac{{\sqrt A }}{2}} \right] $ by the surface region of the $\left( {{n_y},{n_z}} \right)$th antenna element, and denote $ {{\mathbf{e}}_x} = {\left[ {1,0,0} \right]^T}$ by the UPA normal vector, i.e., the unit vector along the $x$ direction.  By taking into account the basic free-space propagations for a theoretical analysis, the channel power gain between the user and the  $ \left( {{n_y},{n_z}} \right) $th antenna element is given by \cite{XLM-8}
\begin{align}\label{channel_power_gain}
&	{f^{\left( {{n_y},{n_z}} \right)}}\left( {{\theta _l},{\varphi _l}, {r_l}} \right) \nonumber\\
	&= \int_{{\mathcal{A}_{{n_y},{n_z}}}} {\underbrace {\frac{1}{{4\pi {{\left\| {{\mathbf{s}} - {\mathbf{\bar p}}} \right\|}^2}}}}_{{\text{Free-space path loss}}}} \underbrace {\frac{{{{\left( {{\mathbf{s}} - {\mathbf{\bar p}}} \right)}^T}{{\mathbf{e}}_x}}}{{\left\| {{\mathbf{s}} - {\mathbf{\bar p}}} \right\|}}}_{{\text{Projection to signal direction}}}d{\mathbf{\bar p}},
\end{align}
where $r_l$ represents the distance between the user (or scatterer) and the reference antenna element of the UPA (i.e., $\left( {{n_y},{n_z}} \right) = \left( {0,0} \right)$), and  $ \left( {{\theta _l},{\varphi _l}} \right) $ denotes the azimuth-elevation angle pair of the $l$th path, with $l=1$ being the line-of-sight (LoS) path and  $ l \ge 2 $ being non-line-of-sight (NLoS) paths. The user location $\bf{s}$ is determined by the LoS path, i.e., ${\mathbf{s}} = {\left[ {{r_l}\sin {\theta _l}\cos {\varphi _l},{r_l}\sin {\theta _l}\sin {\varphi _l},{r_l}\cos {\theta _l}} \right]^T},l = 1$. The examination in (\ref{channel_power_gain}) is applied to each individual antenna element, taking explicit account of the element size. More explicitly, the model in (\ref{channel_power_gain}) accounts for the projected aperture of each antenna element, as reflected by the projection of the UPA normal vector ${{\mathbf{e}}_x}$ to the wave propagation direction at each local point ${\mathbf{\bar p}}$, in contrast to the conventional model for free-space path loss.
In \cite{XLM_8-11,XLM_8-18,XLM_8-19}, similar models incorporating the projected aperture of the antenna element have also been considered for communicating with continuous surfaces. Hence, a two-dimensional integration over the surface of each element is required for a precise evaluation of (\ref{channel_power_gain}).

Owing to the fact that each antenna element (not the entire array) in $A$ is on the wavelength scale, there may be typically marginal variations in the wave propagation distance $\left\| {{\mathbf{s}} - {\mathbf{\bar p}}} \right\|$ and projection direction $\frac{{{\mathbf{s}} - {\mathbf{\bar p}}}}{{\left\| {{\mathbf{s}} - {\mathbf{\bar p}}} \right\|}}$ across different points ${\mathbf{\bar p}} \in {\mathcal{A}_{{n_y},{n_z}}}$. Hence, in light of the fact in \cite{XLM-8}, we have
\begin{align}
&\frac{1}{{4\pi {{\left\| {{\mathbf{s}} - {\mathbf{\bar p}}} \right\|}^2}}} \approx \frac{1}{{4\pi {{\left\| {{\mathbf{s}} - {{\mathbf{p}}_{{n_y},{n_z}}}} \right\|}^2}}}, \nonumber\\
&\frac{{{{\left( {{\mathbf{s}} - {\mathbf{\bar p}}} \right)}^T}{{\mathbf{e}}_x}}}{{\left\| {{\mathbf{s}} - {\mathbf{\bar p}}} \right\|}} \approx \frac{{{{\left( {{\mathbf{s}} - {{\mathbf{p}}_{{n_y},{n_z}}}} \right)}^T}{{\mathbf{e}}_x}}}{{\left\| {{\mathbf{s}} - {{\mathbf{p}}_{{n_y},{n_z}}}} \right\|}},\forall {\mathbf{\bar p}} \in {\mathcal{A}_{{n_y},{n_z}}},
\end{align}
which facilitates the approximation of the channel power gain in (\ref{channel_power_gain}), i.e.,
\begin{align}
{f^{\left( {{n_y},{n_z}} \right)}}\left( {{\theta _l},{\varphi _l}, {r_l}} \right) &\approx \frac{1}{{4\pi {{\left\| {{\mathbf{s}} - {{\mathbf{p}}_{{n_y},{n_z}}}} \right\|}^2}}}\underbrace {A\frac{{{{\left( {{\mathbf{s}} - {{\mathbf{p}}_{{n_y},{n_z}}}} \right)}^T}{{\mathbf{e}}_x}}}{{\left\| {{\mathbf{s}} - {{\mathbf{p}}_{{n_y},{n_z}}}} \right\|}}}_{{\text{Projected aperture}}} \nonumber\\
&= \frac{{A{r_l}\sin {\theta _l}\cos {\varphi _l}}}{{4\pi {{\left\| {{\mathbf{s}} - {{\mathbf{p}}_{{n_y},{n_z}}}} \right\|}^3}}}.
\end{align}
Accordingly, the holographic channel ${\mathbf{h}}$ between the BS and the user can be modeled~as~\cite{Holo-15}
\begin{equation}
	{\mathbf{h}} = \sqrt {\frac{1}{L}} \sum\limits_{l = 1}^L {{\beta _l}{\mathbf{a}}\left( {{\theta _l},{\varphi _l},{r_l}} \right)} ,
\end{equation}
where $\beta_l$ denotes the channel coefficient of the $l$th path and $L$ is the path number. Note that for the LoS path, i.e., $l=1$, we denote $\beta_l = \sqrt {{f^{\left( {{n_y},{n_z}} \right)}}\left( {{\theta _l},{\varphi _l}, {r_l}} \right)}$, since the LoS path is distance-determinate~\cite{Holo-15,chen-jsac}. For the NLoS path, i.e., $l \ge 2$, the channel coefficient $\beta_l$ follows a complex Gaussian distribution, i.e., $\beta_{l} \sim \mathcal{CN} \left(  { 0, {f^{\left( {{n_y},{n_z}} \right)}}\left( {{\theta _l},{\varphi _l}, {r_l}} \right) } \right) $.
The array response vector ${\mathbf{a}}\left( {{\theta _l},{\varphi _l},{r_l}} \right) \in {\mathbb{C}^{N \times 1}}$ is structured by the following entry 
\begin{equation}
{a^{\left( {{n_y},{n_z}} \right)}}\left( {{\theta _l},{\varphi _l},{r_l}} \right) =  \exp \left\{ { - \jmath\frac{{2\pi }}{\lambda }\left( {r_l^{\left( {{n_y},{n_z}} \right)} - {r_l}} \right)} \right\},
\end{equation}
where $ \Delta r_l^{\left( {{n_y},{n_z}} \right)} \triangleq r_l^{\left( {{n_y},{n_z}} \right)} - {r_l} $ represents the extra-distance traveled by the wave to arrive at the  $ \left( {n_y, n_z} \right)  $th antenna element with respect to the reference one, and it can be unfolded as
\begin{equation}\label{delta_r}
\resizebox{\hsize}{!}{$
\Delta r_l^{\left( {{n_y},{n_z}} \right)} = {r_l}\left[ {{{\left( {1 + {{\left( {\frac{{{r^{\left( {{n_y},{n_z}} \right)}}}}{{{r_l}}}} \right)}^2} - 2\frac{{{r^{\left( {{n_y},{n_z}} \right)}}}}{{{r_l}}}g_l^{\left( {{n_y},{n_z}} \right)}} \right)}^{1/2}} - 1} \right], $}
\end{equation}
where $ {r^{\left( {{n_y},{n_z}} \right)}} \triangleq \delta \sqrt {\left( {n_y^2 + n_z^2} \right)}  $ herein represents the distance between the $ \left( {{n_y},{n_z}} \right) $th antenna element and the reference one, and $g_l^{\left( {{n_y},{n_z}} \right)} = \cos \vartheta ^{\left( {{n_y},{n_z}} \right)}\sin {\theta _l}\sin {\varphi _l} + \sin \vartheta ^{\left( {{n_y},{n_z}} \right)}\cos {\theta _l} = \frac{1}{{\sqrt {n_y^2 + n_z^2} }}   \left( {  {n_y}\sin {\theta _l}\sin {\varphi _l} + {n_z}\cos {\theta _l}} \right)$ is a geometry term, with $\vartheta ^{\left( {{n_y},{n_z}} \right)} = \arctan \frac{{{n_y}}}{{{n_z}}}$ indicating the azimuth angle of the $\left( {{n_z},{n_y}} \right)$th antenna element with respect to the reference one.


For uplink channel estimation, it is assumed that the users transmit mutual orthogonal pilot sequences to the BS, e.g., by employing orthogonal frequency or time resources for different users, hence allowing the channel estimation to be independent of one another. Without loss of generality, we examine the channel estimation process of an arbitrary user. Denote $P$ by the pilot number and $x_p$ by user's information-bearing signal, respectively, and the received signal ${{\mathbf{y}}_p} \in {\mathbb{C}^{{N_{{\text{RF}}}} \times 1}}$ at the BS is given by
\begin{equation}\label{received_signal}
{{\mathbf{y}}_p} = {{\mathbf{F}}_p}{{\mathbf{M}}_p}{\mathbf{h}}{x_p} + {{\mathbf{F}}_p}{{\mathbf{M}}_p}{{\mathbf{n}}_p},p = 1,...,P,
\end{equation}
where ${{\mathbf{F}}_p} \in {\mathbb{C}^{{N_{{\text{RF}}}} \times N}}$ represents the combining matrix at the BS, ${{\mathbf{M}}_p} \in {\mathbb{C}^{N \times N}}$ is the holographic pattern that records the radiation amplitude of each antenna element, ${{\mathbf{n}}_p} \in {\mathbb{C}^{N \times 1}} \sim \mathcal{C}\mathcal{N}\left( {0,{\sigma ^2}{{\mathbf{I}}_N}} \right)$ denotes the additive white Gaussian noise (AWGN) with power $\sigma^2$, and ${\mathbf{h}} \in {\mathbb{C}^{N \times 1}}$ characterizes the channel spanning from the BS to the user. The aggregated received pilot sequence ${\mathbf{y}} = {\left[ {{\mathbf{y}}_1^T,...,{\mathbf{y}}_P^T} \right]^T} \in {\mathbb{C}^{P{N_{{\text{RF}}}} \times 1}}$ of all the $P$ uplink pilots from the user is given by
\begin{equation}
{\mathbf{y}} = {\mathbf{Fh}} + {\mathbf{n}},
\end{equation}
in which ${\mathbf{n}} = {\left[ {{\mathbf{n}}_1^T{{\left( {{{\mathbf{F}}_1}{{\mathbf{M}}_1}} \right)}^T},...,{\mathbf{n}}_P^T{{\left( {{{\mathbf{F}}_P}{{\mathbf{M}}_P}} \right)}^T}} \right]^T} \in {\mathbb{C}^{P{N_{{\text{RF}}}} \times 1}}$ is the noise matrix, and $ {\mathbf{F}} = \left[ {{{\left( {{{\mathbf{F}}_1}{{\mathbf{M}}_1}} \right)}^T}x_1,...,  {{\left( {{{\mathbf{F}}_P}{{\mathbf{M}}_P}} \right)}^T}x_P} \right]^T \in {\mathbb{C}^{P{N_{{\text{RF}}}} \times N}} $ delivers the overall observation matrix. 

\begin{figure}
	\subfigure[]{
		\begin{minipage}[t]{0.45\textwidth}
			\centering
			\includegraphics[width=\textwidth]{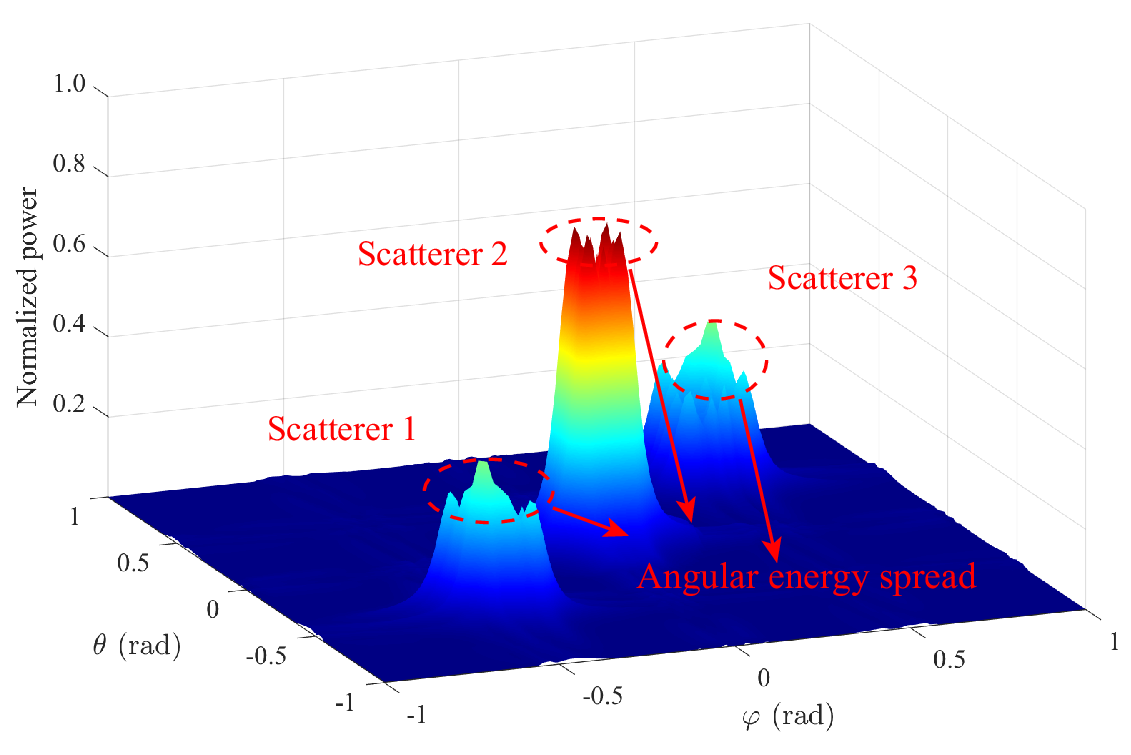}
			\label{3d_angular_spread}
	\end{minipage}}
	\subfigure[]{
		\begin{minipage}[t]{0.45\textwidth}
			\centering
			\includegraphics[width=\textwidth]{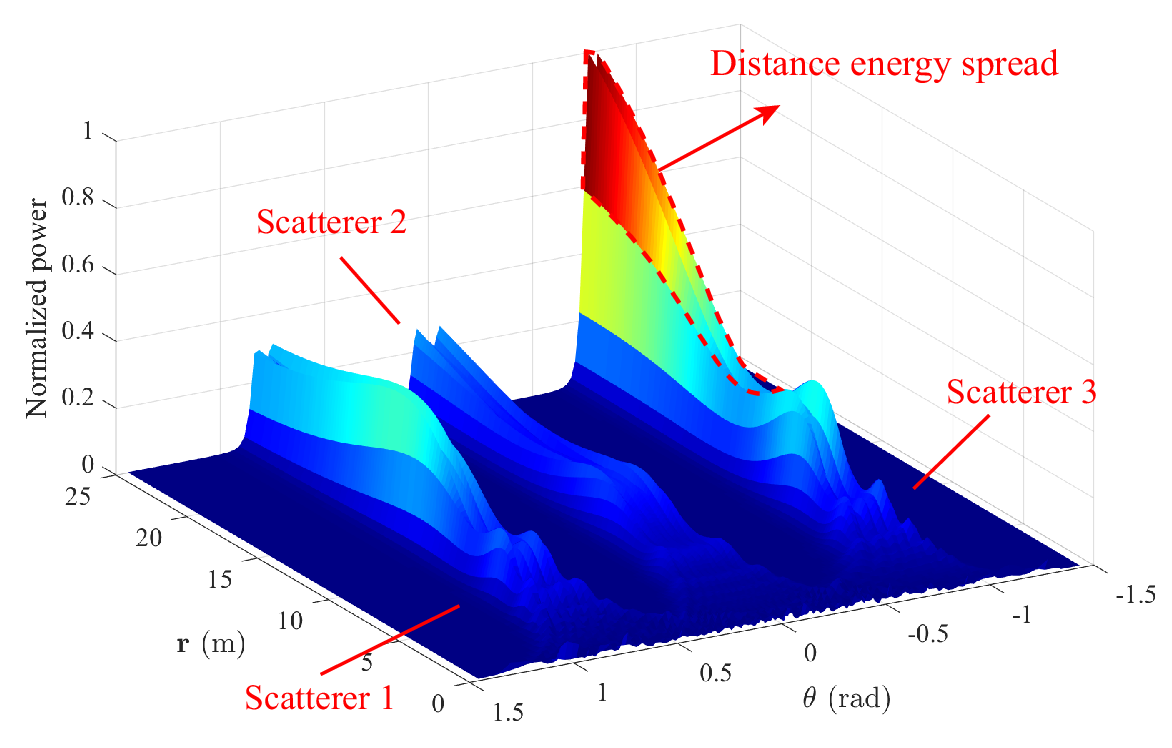}
			\label{3d_distance_spread}
	\end{minipage}}
	\caption{(a) Angular energy spread: the normalized power scattered results in the proliferation of a multitude angles. (b) Distance energy spread: the normalized power scattered in relation to the distance experiences a fluctuating from peak to tough.}\label{challenge}
\end{figure}

\subsection{Challenges}
Most channel estimation schemes, being based on the classical planar-wavefront assumption, exploit the sparsity in the angular domain to achieve efficient estimation. However, this may result in severe energy spread effect in holographic MIMO counterparts. To be explicit, 
Fig.~\ref{challenge} portrays the power scattered in relation to the angle pair ($\bm{\theta}, \bm{\varphi}$) in Fig.~\ref{3d_angular_spread}, and the power scattered in relation to the distance $\bf{r}$ in Fig.~\ref{3d_distance_spread}, respectively, in which three scatterers are assumed in the propagation environment. 
As transpired in Fig.~\ref{3d_angular_spread}, the normalized power scattered typically results in the proliferation of angles, 
and fluctuates rapidly across the holographic antenna array. Despite the detachment of distance information $\bf{r}$, the true azimuth-elevation angle pair $(\bm{\theta}, \bm{\varphi})$ cannot be precisely distinguished. Additionally, as portrayed in Fig.~\ref{3d_distance_spread}, the power scattered versus the distance reveals a fluctuating trend from peak to trough. In this case, it is difficult to distinguish the genuine scattered power of the exact distance, since the energy may spread farther away with increased distance. 


Actually, the angular-domain transform matrix just samples the angular information but ignores the distance information, which is implausible for the holographic near-field context due to the angles being functioned by the distance. Thus, the angular-domain transform matrix fails to align the sampling points to the genuine scattered power of the significant path corresponding to a user/scatterer.
This always leads to a false alarm, i.e., the genuine channel power corresponding to the non-significant path is misinterpreted as a significant one, thus eroding the detection performance of the 3D AED parameters. To tackle this issue, we intend to achieve the effect decomposition of the 3D AED parameters by constructing their respective covariance matrices, for which each covariance matrix constructed retains its self-contained information while canceling out the potential impact of other parameters.

\section{Decomposition of 3D AED Parameters $ \bm{\theta}$, $\bm{\varphi}$, and $\bf{r}  $}

Given that the bare minimum of channel structure can be used when doing channel estimation, let us define $ {\mathbf{V}} \in {\mathbb{C}^{N_y \times N_z}}  $ as the matrix involving some appropriately selected entries from the covariance matrix of the exact channel, whose $\left( {{n_y},{n_z}} \right)$th entry can be given by
\begin{align}\label{v_n}
&{v_{{n_y},{n_z}}}  \triangleq \mathbb{E}\left\{ {{h_{{n_y},{n_z}}}h_{{n^\prime_y},{n^\prime_ z}}^*} \right\} \nonumber\\ &
 \resizebox{\hsize}{!}{$= \sum\limits_{l = 1}^L {\sum\limits_{l' = 1}^L {\mathbb{E}\left\{ {{\beta _l}\beta _{l'}^*} \right\}\exp \left( {\jmath \frac{{2\pi }}{\lambda }\Delta r_{l'}^{ \left(  {n^\prime_y},{n^\prime_z} \right) } - \jmath \frac{{2\pi }}{\lambda }\Delta r_l^{ \left( {n_y},{n_z} \right)  }} \right)} } ,\forall l \ne l' ,$} \nonumber\\
& \mathop  = \limits^{\left( a \right)} \sum\limits_{l = 1}^L \bar \beta _l^2 \exp \left( {\jmath\frac{{2\pi }}{\lambda }\left( {\Delta r_l^{\left(  {n ^\prime_y},{n^\prime_z} \right) } - \Delta r_l^{ \left( {n_y},{n_z} \right)  } } \right)} \right) ,
\end{align}
where $\bar \beta _l^2 \triangleq \mathbb{E}\left\{ {\beta _l^2} \right\}$ and $\left( a \right) $ holds owing to the orthogonal assumption between two independent paths. Although $\bf{V}$ represents a part of the covariance matrix, we refer to it as the covariance matrix throughout of this paper. As observed in (\ref{v_n}),  $v_{n_y,n_z}$ is dependent on the distance difference ${\Delta r_l^{ \left( n ^\prime _y, n^\prime _z \right)  } } - \Delta r_l^{ \left(  {n_y},{n_z} \right)  }$ with respect to the antenna elements $(n_y,n_z)$ and $(n^\prime_y,n^\prime_z)$. Using the Fresnel approximation \cite{Fresnel}, (\ref{delta_r}) can be further recast to
\begin{equation}\label{Fresnel}
	\resizebox{\hsize}{!}{$
\Delta r_l^{\left( {{n_y},{n_z}} \right)} \approx \frac{{\left( {n_y^2 + n_z^2} \right){\delta ^2}}}{{2{r_l}}}\left( {1 - {{\left( {g_l^{\left( {{n_y},{n_z}} \right)}} \right)}^2}} \right) - \sqrt {n_y^2 + n_z^2} \delta g_l^{\left( {{n_y},{n_z}} \right)} $} ,
\end{equation}
in which the first term involves both the angular and distance information, while the second term pertains exclusively to the angular information, i.e., $g_l ^ {\left( {n_y,n_z} \right)} $. This property is paramount since it inspires the subsequent parametric deconstruction. In the subsections that follows, tailored tricks are presented for effectively decomposing the tightly coupled 3D AED parameters.

\begin{figure*}[t]
	\centering
	\includegraphics[width=0.85\textwidth]{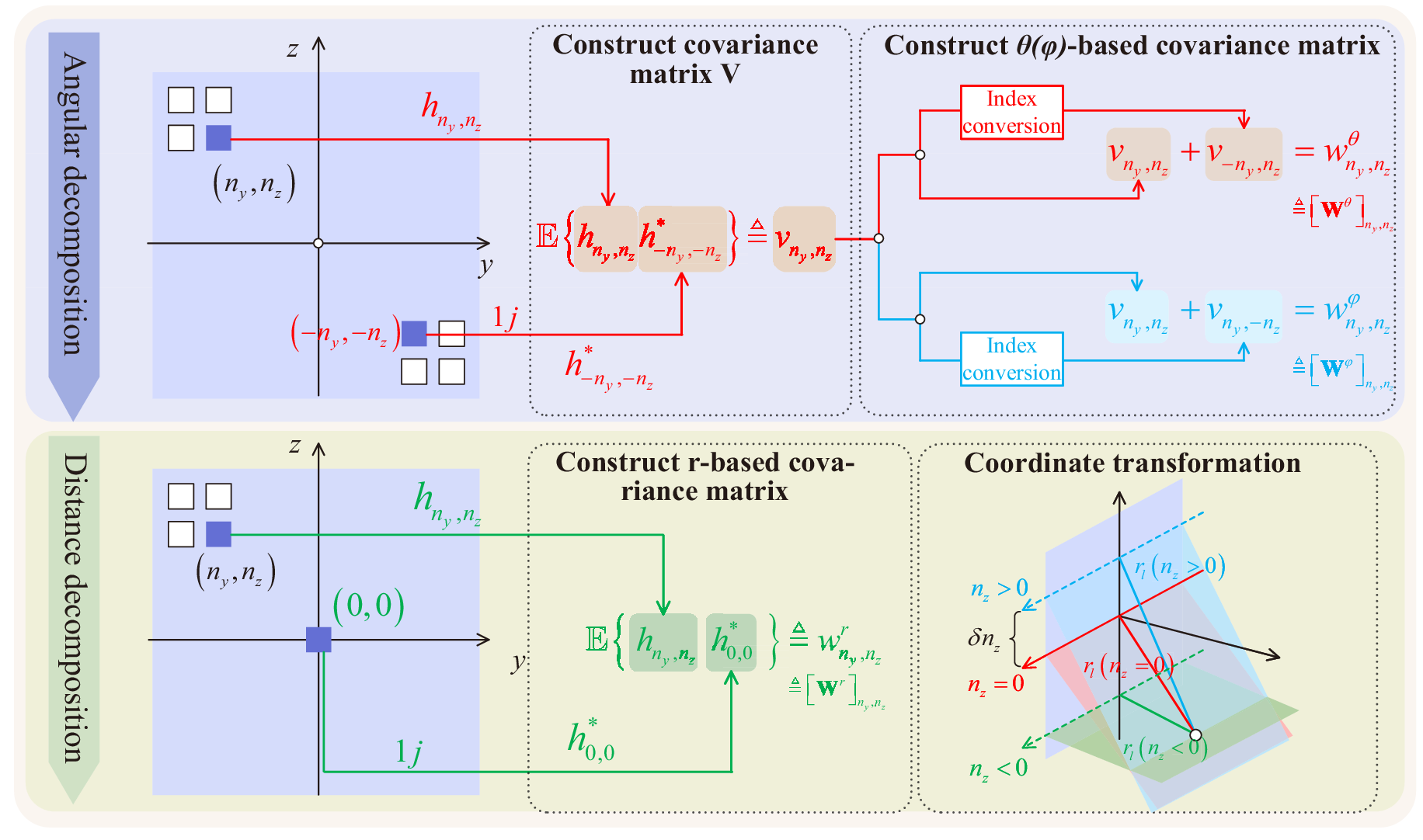}
	\caption{Schematic illustration of parametric decomposition.} \label{DeRe}
\end{figure*}

\subsection{Angular Decomposition}

The structure of (\ref{Fresnel}) inspires a potential technique to select the index of the ${\left(  {n ^\prime_y},{n^\prime_z} \right) }$th entry of the conjugate channel $h_{  {n ^\prime_y},{n^\prime_z}  }^*$ with extra care for ensuring that $v_{n_y,n_z}$  is determined exclusively by the angular information. The index selection strategy 
needs to satisfy $ g_l^{\left( {n^\prime_y,n^\prime_z} \right)} =  - g_l^{\left( {{n_y},{n_z}} \right)} $ and $  {\left( {{n ^ \prime _y}} \right)^2} + {\left( {{n^\prime _z}} \right)^2}=n_y^2 + n_z^2  $. This pair of restrictions can be readily addressed by letting $ \left( {{n^\prime_y},{n^\prime_z}} \right) =  \left( { - {n_y}, - {n_z}} \right) $, and thus $v_{n_y,n_z}$ in (\ref{v_n}) can  be trimmed to
\begin{align}\label{v_n_angle}
&{v_{{n_y},{n_z}}} \left( {\theta_l,\varphi_l} \right) \nonumber\\
 &= \sum\limits_{l = 1}^L  \bar \beta _l^2 \exp \left( {\jmath\frac{{4\pi }}{\lambda }\sqrt {n_y^2 + n_z^2} \delta g_l^{\left( n_y, n_z \right)}} \right)  \nonumber\\
&= \sum\limits_{l = 1}^L \bar \beta _l^2 \exp \left( {\jmath\frac{{4\pi }}{\lambda }\delta \left( {{n_y}\sin {\theta _l}\sin {\varphi _l} + {n_z}\cos {\theta _l}} \right)} \right).
\end{align}
The physical significance behind this index selection trick is to allow the indices of the antenna elements corresponding to the channel $h_{n_y,n_z}$ and its conjugate copy $h_{n^\prime_y,n^\prime_z}^*$ to be symmetrical to the origin, as illustrated in Fig.~\ref{DeRe}. As a result, $ {v_{{n_y},{n_z}}} \left( {\theta_l,\varphi_l} \right) $ retains only the angular information of the original channel structure.
To facilitate the reduced searching space in the estimation process, the decomposition can be further refined to trim the 2D model at hand into a pair of 1D quantities consisting exclusively of $\bm{\theta}$ or $\bm{\varphi}$.

\subsubsection{Extraction of $\bm{\theta}$}
We aim to retain the $\bm{\theta}$-associated term in the phase of $v_{n_y,n_z} \left( {\theta_l,\varphi_l} \right) $, i.e., $n_z \cos \theta_l$, by constructing a $\bm{\theta}$-based covariance matrix $ {{\mathbf{W}}^\theta } $. Particularly, the $\left( n_y,n_z\right) $th entry in $ {{\mathbf{W}}^\theta } $ can be calculated by $ w_{{n_y},{n_z}}^\theta  = v_{{n_y},{n_z}} \left( {{\theta _l},{\varphi _l}} \right) + v_{n^\prime_y,n^\prime_z} \left( {{\theta _l},{\varphi _l}} \right) $. By letting $ \left( {n^\prime_y},{n^\prime_z} \right) = \left( { - {n_y},{n_z}} \right) $, we~have
\begin{align}\label{tmp-12}
&w_{{n_y},{n_z}}^\theta = {v_{{n_y},{n_z}}}\left( {{\theta _l},{\varphi _l}} \right) + {v_{ - {n_y},{n_z}}}\left( {{\theta _l},{\varphi _l}} \right) \nonumber\\
& = \sum\limits_{l = 1}^L {\bar \beta _l^2}\exp \left( {\jmath\frac{{4\pi }}{\lambda }\delta \left( {{n_y}\sin {\theta _l}\sin {\varphi _l} + {n_z}\cos {\theta _l}} \right)} \right) \nonumber\\
& \quad + \sum\limits_{l = 1}^L {\bar \beta _l^2}\exp \left( {\jmath \frac{{4\pi }}{\lambda }\delta \left( { - {n_y}\sin {\theta _l}\sin {\varphi _l} + {n_z}\cos {\theta _l}} \right)} \right)  \nonumber\\
& = \sum\limits_{l = 1}^L 2{\bar \beta _l^2} \cos \left( {\frac{{4\pi }}{\lambda }\delta {n_y}\sin {\theta _l}\sin {\varphi _l}} \right)\exp \left(  \jmath \frac{{4\pi }}{\lambda }\delta {n_z}\cos {\theta _l} \right) .
\end{align}
We let the antenna element spacing be $ \delta  = \frac{\lambda }{4} $, and the $\left( n_y, n_z \right) $th entry of the $\bm{\theta}$-based covariance matrix $ {{\mathbf{W}}^\theta } $ takes the form of
\begin{equation}\label{w_theta}
w_{{n_y},{n_z}}^\theta = \sum\limits_{l = 1}^L {2  {\bar \beta _l^2}  } \cos \left( {\pi {n_y}\sin {\theta _l}\sin {\varphi _l}} \right)\exp \left( {\jmath \pi {n_z}\cos {\theta _l}} \right).
\end{equation}
Note that $ \delta  = \frac{\lambda }{4} $ conveys double implications, that is, it not only indicates a denser antenna spacing in the holographic MIMO of interest, but it also complies with the Nyquist sampling theorem under the DeRe framework for acquiring  statistical characteristics of $\bm{\theta}$ (i.e., $\bm{\theta}$-based covariance matrix ${\bf{W}} ^ \theta$). To be explicit,
it is evident from (\ref{tmp-12}) that the estimation of $\bm{\theta}$ is equivalent to identifying all $L$ spectral power peaks with the frequency index $ \left\lbrace {\cos \theta_l, l =1,...,L} \right\rbrace  $ given the sequences $ w_{{n_y},{n_z}}^\theta  $ with the time index~$n_z$. The phase term in (\ref{tmp-12}), i.e., $\exp \left(  \jmath \frac{{4\pi }}{\lambda }\delta {n_z}\cos {\theta _l} \right)  $, can be structured as a standard Fourier-like form, namely $\exp \left(  \jmath 2 \pi n_z \cdot \left( {\frac{2\delta }{\lambda}\cos \theta_l} \right)  \right)  $. The frequency component $ \frac{2\delta }{\lambda}\cos \theta_l $ has to follows Nyquist sampling theorem, yielding that $ \left| {\frac{{2\delta }}{\lambda }\cos {\theta _l}} \right| \le \frac{{2\delta }}{\lambda } \le \frac{1}{2} $, and we thus have $ \delta  \le \frac{\lambda }{4} $. Additionally, as for the constructed $\bm{\theta}$-based covariance matrix ${\bf{W}}^\theta \in \mathbb{C}^{N_y \times N_z} $, the row vectors thereof, i.e., $  {\left[ {{{\mathbf{W}}^\theta }} \right]_{{n_y},:}} $, are independent of one another. Each independent vector delivers an observation indexed by $n_y$ for the estimate of $\bm{\theta} $, i.e., $ {{\hat \theta }_{{n_y}}} $, in which these $N_y$  estimates $ {{\hat \theta }_{{n_y}}},{n_y} = 1,...,{N_y}, $ contribute in collaboration to a faithful $\hat{ \bm{\theta}}$, as elaborated on later.

\subsubsection{Extraction of $\bm{\varphi}$}
The same technique can be applied to construct the $\bm{\varphi}$-based covariance matrix ${\bf{W}}^\varphi$, but using a different index-selection strategy, i.e., $ \left( {{n^\prime_y},{n^\prime_z}} \right) = \left( {{n_y}, - {n_z}} \right) $, hence~yielding
\begin{align}\label{w_varphi}
{w^\varphi_{{n_y},{n_z}}} &= {v_{{n_y},{n_z}}}\left( {{\theta _l},{\varphi _l}} \right) + {v_{{n^\prime_y},{n^\prime_z}}}\left( {{\theta _l},{\varphi _l}} \right) \nonumber\\
 &= \sum\limits_{l = 1}^L {2\bar \beta _l^2} \cos \left( {\pi {n_z}\cos {{ \theta }_l}} \right)\exp \left( {\jmath \pi {n_y}\sin {{ \theta }_l}\sin {\varphi _l}} \right).
\end{align}
As shown in (\ref{w_varphi}), both the amplitude and phase of $\bm{\varphi}$ contains the information of $ {\bm{\theta}}$, which implies that an estimate of  $\hat {\bm{\theta}}$ is a prerequisite for the determination of $\bm{\varphi}$. The $\bm{ \varphi }$-based covariance matrix ${\bf{W}}^\varphi \in \mathbb{C} ^{N_y \times N_z}$ incorporates $N_z$ independent vectors with vector size~$N_y$. In other words, the $n_z$th column vector of ${\bf{W}}^\varphi $ serves as an observation that can be used for the CS process to attain $\hat{\varphi} _ {n_z}, n_z = 1,..., N_z$, the $N_z$ estimates of which can be appropriately fused by a sophisticated design for the robust recovery of ${\bm{\varphi}}$. Further details appear in Sec.~V.


\subsection{Distance Decomposition}

The distance decomposition can be achieved by constructing
the $ \bf{r} $-based covariance matrix ${\bf{W}}^r$ that strips out the angular information for retrieving the distance information. Specifically, we use the same trick when constructing ${\bf{W}}^r$. As illustrated in Fig.~\ref{DeRe}, we employ the channel $h_{n_y,n_z}$ and its conjugate copy corresponding to the origin of the
UPA, namely $h_{0,0} ^*$, as a pair of inputs for $w^r_{n_y,n_z}$, i.e.,
\begin{align}\label{w_r}
w_{{n_y},{n_z}}^r & \triangleq \mathbb{E}\left\{ {{h_{{n_y},{n_z}}}h_{0,0}^*} \right\} \nonumber\\
&= \sum\limits_{l = 1}^L {\bar \beta _l^2\exp \left( {\jmath \frac{{2\pi }}{\lambda }\left( {\Delta r_l^{\left( {{n_y},{n_z}} \right)} - \Delta r_l^{\left( {0,0} \right)}} \right)} \right)} .
\end{align}
Recall that the distance difference takes the form of $\Delta r_l^{\left( {{n_y},{n_z}} \right)} = \frac{{\left( {n_y^2 + n_z^2} \right){\delta ^2}}}{{2{r_l}}}\left( {1 - {{\left( {g_l^{\left( {{n_y},{n_z}} \right)}} \right)}^2}} \right) - \delta \times \\ \left( {{n_y}\sin {\theta _l}\sin {\varphi _l} + {n_z}\cos {\theta _l}} \right)$, with $g_l^{\left( {{n_y},{n_z}} \right)} = \cos \vartheta ^{\left( {{n_y},{n_z}} \right)}\sin {\theta _l}\sin {\varphi _l}  + \sin \vartheta ^{\left( {{n_y},{n_z}} \right)}\cos {\theta _l} =  \frac{1}{{\sqrt {n_y^2 + n_z^2} }} \\ \times  \left( {  {n_y}\sin {\theta _l}\sin {\varphi _l} + {n_z}\cos {\theta _l}} \right) $. By substituting them into (\ref{w_r}), we have (\ref{w_r_1}) presented at the top of the next page.
\begin{figure*}
\begin{equation}\label{w_r_1}
w_{{n_y},{n_z}}^r  = \sum\limits_{l = 1}^L {\bar \beta _l^2} \exp \left( {\frac{{\jmath \pi }}{4}\left[ {\frac{{\left( {n_y^2 + n_z^2} \right)\delta }}{{{r_l}}}\left( {1 - {{\left( {g_l^{\left( {{n_y},{n_z}} \right)}} \right)}^2}} \right) - 2\left( {{n_y}\sin {\theta _l}\sin {\varphi _l} + {n_z}\cos {\theta _l}} \right)} \right]} \right).
\end{equation}
\hrule
\end{figure*}
For simplicity, we firstly consider a special case of $n_z = 0$, and hence (\ref{w_r_1}) can be trimmed to (\ref{w_r_2}), shown at the top of the next page.
\begin{figure*}
\begin{equation}\label{w_r_2}
w_{{n_y},{n_z}}^r = \sum\limits_{l = 1}^L {\bar \beta _l^2} \exp \left( {\jmath \left[ {\frac{{\pi n_y^2\lambda }}{{16{r_l}}}\left( {1 - {{\left( {g_l^{\left( {{n_y},{n_z}} \right)}} \right)}^2}} \right) - \frac{\pi }{2}{n_y}\sin {\theta _l}\sin {\varphi _l}} \right]} \right).
\end{equation}
\hrule
\end{figure*}
A closed examination in (\ref{w_r_2}) indicates that the recovery of the distance $\bf{r}$ from ${\bf{W}}^r$ relies upon the acquisition of the azimuth-elevation angle pair $ \left( {\bm{\theta}, \bm{\varphi}} \right)  $, which has a direct impact on the algorithm flow, as will be detailed in Sec.~V. Additionally, for a more general instance of $n_z \neq 0$, the same form as (\ref{w_r_2}) can be attained by shifting the reference system by $ \delta n_z $ (m) along the $z$-axis using the coordinate transformation.

Thus far, we have achieved the decomposition of the 3D AED parameters $\left( {\bm{\theta}, \bm{\varphi}, \bf{r}} \right) $ by constructing their associated covariance matrices, each of which retains fully knowledge that can be explored. In light of the closed structures in (\ref{w_theta}), (\ref{w_varphi}) and (\ref{w_r_2}), we are inspired to conduct the CS procedure on each parameter to ensure their accurate and robust recoveries. Therefore, the following of this treatise moves on to describe in great detail the CS-based reconstruction, along with an efficient algorithm proposed for resolving CS-related setbacks.

\section{CS-based Reconstruction}


\subsection{CS Formulation of $\bm{\theta}$}

The constructed $\bm{\theta}$-based covariance matrix $ {{\mathbf{W}}^\theta } \in \mathbb{C}^{N_y \times N_z} $ encompasses $N_y$ vectors with vector size $N_z$, in which the $n_y$th vector can be drawn out from $ {\mathbf{W}}^\theta $ given by $ { \mathbf{w}}_{{n_y}}^\theta  \triangleq {\left( {{{\left[ {{{\mathbf{W}}^\theta }} \right]}_{{n_y},:}}} \right)^T} \in {\mathbb{C}^{{N_z} \times 1}} $. The vector $ {\mathbf{w}}_{{n_y}}^\theta $ is actually an observation interpreted as a linear combination of all
$ L $ steering vectors shown in (\ref{w_theta}). Furthermore, a wide measurement matrix $ {\mathbf{B}}_{{n_y}}^\theta  \in {\mathbb{C}^{{N_{ \text{RF}  }} \times {N_z}}}\left( {{N_{ \text{RF} }} \ll {N_z}} \right) $ can be introduced for a lower-dimension observation ${\mathbf{y}}_{{n_y}}^\theta $, thus yielding
\begin{equation}\label{theta_CS_1}
{\mathbf{y}}_{{n_y}}^\theta  = {\mathbf{B}}_{{n_y}}^\theta {\mathbf{w}}_{{n_y}}^\theta ,
\end{equation}
where the entries in $ {\mathbf{B}}_{{n_y}}^\theta $ are allowed to follow a complex Gaussian distribution $ \mathcal{C}\mathcal{N}\left( {0,1/\sqrt{N_y}} \right) $.

Given that there are only $ L $ significant angles of arrival (AoAs) $ \bm{\theta} = \left\{ {{\theta _1},...,{\theta _L}} \right\} $ in the physical propagation environment, it is beneficial to mine the sparse characteristics concealed behind ${\mathbf{w}}_{{n_y}}^\theta$ for achieving sparse estimates of $ L $ AoAs at reduced complexity and overhead. For this purpose, a uniform grid of $ {\tilde N_z} $ AoAs $ \bm{\theta}  $  over $ \left[ { - \pi/2, \pi/2} \right)  $ are prescribed to take values from the discrete sets $ \left\{ {{{\tilde \theta }_{{n_z}}}:\sin {{\tilde \theta }_{{n_z}}} = \frac{2}{{{{\tilde N}_z}}}\left( {{n_z} - \frac{{{{\tilde N}_z} - 1}}{2}} \right),{n_z} = 0,...,{{\tilde N}_z} - 1} \right\} $. Nevertheless, the true AoAs do not coincide exactly with the grid points. To capture the attributes of mismatches between the true angles and the grid points, the off-grid basis is harnessed for its sparse representation \cite{chen-twc3}. More precisely, we introduce the $ \bm{\theta} $-based offset vector $ \Delta \bm{\theta } = \left\{ {\Delta {\theta _1},...,\Delta {\theta _{{{\tilde N}_z}}}} \right\} $, in which $ \Delta {\theta _{{n_z}}} $ is given by
\begin{equation}
\Delta {\theta _{{n_z}}} = \left\{ \begin{gathered}
	{\theta _{{n_z}}} - {{\tilde \theta }_{{n_{z,q}}}},  \text{if} \ {n_z} = {n_{z,q}},q = 1,...L, \hfill \\
	0, \qquad \quad \ \  {\text{otherwise}}, \hfill \\ 
\end{gathered}  \right.
\end{equation}
in which $n_{z,q}$ indicates the index of the AoA grid point nearest to the AoA of the $l$th path. Thus, $ {\mathbf{w}}_{{n_y}}^\theta  $ can be structured as 
\begin{equation}\label{w_CS_theta}
{\mathbf{w}}_{{n_y}}^\theta  = {\mathbf{\Psi }}_{{n_y}}^\theta \left( {\Delta \bm{\theta }} \right){\mathbf{u}}_{{n_y}}^\theta  + {\mathbf{n}}_{{n_y}}^\theta ,
\end{equation}
where $ {\mathbf{u}}_{{n_y}}^\theta  \in {\mathbb{C}^{{{\tilde N}_z} \times 1}} $ is an $L$-sparse vector, ${\bf{n}}^\theta _ {n_z} \in {\mathbb{C}^{{{\tilde N}_z} \times 1}} $ is the noise vector, and  $ {\mathbf{\Psi }}_{{n_y}}^\theta  \in {\mathbb{C}^{{N_z} \times {{\tilde N}_z}}} $ is the dictionary matrix. The design of the dictionary matrix is required to satisfy the RIP condition in order to facilitate the accurate sparse signal recovery. As a necessity for accurate sparse signal recovery, the dictionary matrix has to satisfy the RIP condition \cite{chen-twc3} for eliminating the grid coherence, which adversely impacts the estimation performance. To be specific, the RIP condition can be interpreted as two arbitrary columns being (approximately) orthogonal, where the $\left( {{{n_z},{n ^ \prime_z}}} \right) $th entry of $ {\mathbf{\Psi }}_{{n_y}}^\theta  \in {\mathbb{C}^{{N_z} \times {{\tilde N}_z}}} $ is crafted by 
$   {\left[ {{\mathbf{\Psi }}_{{n_y}}^\theta \left( {\Delta \bm{\theta }} \right)} \right]_{\left( {{{n_z},{n ^ \prime_z}}} \right)}} = \exp \left\{ {\jmath \pi {n_z}\cos \left( {{{\tilde \theta }_{{n_y}}} + \Delta {\theta _{{n_y}}}} \right)} \right\}  $. 
Thus, (\ref{theta_CS_1}) can be recast to a CS form with uncertain dictionary matrix (containing the to-be-determined $\Delta \bm{\theta}$) and sparse vector $ {\mathbf{u}}_{{n_y}}^\theta $, i.e.,
\begin{equation}\label{theta_CS_2}
	{\mathbf{y}}_{{n_y}}^\theta  = {\mathbf{B}}_{{n_y}}^\theta {\mathbf{\Psi }}_{{n_y}}^\theta \left( {\Delta \bm{\theta }} \right){\mathbf{u}}_{{n_y}}^\theta  + {\mathbf{B}}_{{n_y}}^\theta {\mathbf{n}}_{{n_y}}^\theta .
\end{equation}

\subsection{CS Formulation of $\bm{\varphi}$}
 
The CS reconstruction for $\bm{\varphi}$ can be executed using similar tactics to those employed in $\bm{ \theta }$'s. However, the phase term in (\ref{w_varphi}) includes the $\bm{\theta}$-associated part, which may preclude the direct gridding over $\bm{\varphi}$. To resolve this hiccup, we introduce an intermediate variable  $ \bm{\zeta } = \left\{ {{\zeta _1},...,{\zeta _L}} \right\} $ for satisfying $\sin {\zeta _l} = \sin {\theta _l}\sin {\varphi _l},l = 1,...,L$. Next, an arbitrary vector indexed by $n_z$ can be extracted from the $\bm{\varphi}$-based covariance matrix ${\bf{W}}^\varphi$, i.e., $ {\mathbf{w}}_{{n_z}}^\zeta  \triangleq {\left[ {{{\mathbf{W}}^\varphi }} \right]_{:,{n_z}}} \in {\mathbb{C}^{{N_y} \times 1}} $. A wide measurement matrix $ {\mathbf{B}}_{{n_z}}^\zeta  \in {\mathbb{C}^{{N_{ \text{RF} }} \times {N_y}}}  $ is left-multiplied by the vector $ {\mathbf{w}}_{{n_z}}^\zeta  $ for confining its size with the number of RF chains, thus yielding ${\mathbf{y}}_{{n_z}}^\zeta  = {\mathbf{B}}_{{n_z}}^\zeta {\mathbf{w}}_{{n_z}}^\zeta $. To perform gridding over $\bm{\zeta}$, a uniform grid of $ {\tilde N_z} $ angles $ \bm{\zeta } $ is introduced, i.e., $ \left\{ {{{\tilde \zeta }_{{n_y}}}:\sin {{\tilde \zeta }_{{n_y}}} = \frac{2}{{{{\tilde N}_y}}}\left( {{n_y} - \frac{{{{\tilde N}_y} - 1}}{2}} \right),{n_y} = 0,...,{{\tilde N}_y} - 1} \right\} $. Regarding the mismatches between the true angles and the prescribed grid points, $\bm{\zeta }$-based offset vector is defined as $ \Delta \bm{\zeta } = \left\{ {\Delta {\zeta _1},...,\Delta {\zeta _{{{\tilde N}_y}}}} \right\} $, with $ \Delta {\zeta _{{n_y}}} $ given by
\begin{equation}
\Delta {\zeta _{{n_y}}} = \left\{ \begin{gathered}
	{\zeta _{{n_y}}} - {{\tilde \zeta }_{{n_{y,q}}}}, \text{if} \ {n_y} = {n_{y,q}},q = 1,...L, \hfill \\
	0, \qquad \quad \ \ {\text{otherwise}}, \hfill \\ 
\end{gathered}  \right.
\end{equation}
in which $ n_{y,q} $  indicates the index of the AoA grid point nearest to the AoA of the $l$th path. Thus, $ {\mathbf{w}}_{{n_z}}^\zeta  $ can be structured as a CS form
\begin{equation}
{\mathbf{w}}_{{n_z}}^\zeta  = {\mathbf{\Psi }}_{{n_z}}^\zeta \left( {\Delta \bm{\zeta }} \right){\mathbf{u}}_{{n_z}}^\zeta  + {\mathbf{n}}_{{n_z}}^\zeta ,
\end{equation}
in which ${\mathbf{u}}_{{n_z}}^\zeta  \in {\mathbb{C}^{{{\tilde N}_y} \times 1}}$ is an $L$-sparse vector, $ {\mathbf{n}}_{{n_z}}^\zeta  \in {\mathbb{C}^{{{\tilde N}_y} \times 1}} $ is the noise vector, and $ {\mathbf{\Psi }}_{{n_z}}^\zeta  \in {\mathbb{C}^{{N_y} \times {{\tilde N}_y}}} $ represents the dictionary matrix, whose $\left( {n_y, n^\prime_y}  \right) $th entry is given by  
{\small$ 	 {\left[ {{\mathbf{\Psi }}_{{n_z}}^\zeta \left( {\Delta \bm{\varphi }} \right)} \right]_{{n_y},{n^\prime_y}}} = \exp \left\{ {\jmath \pi {n_z}\sin \left( {{{\tilde \zeta }_{{n_{z,q}}}} + \Delta {\zeta _{{n_z}}}} \right)} \right\}  $}.
Therefore, the $\bm{\zeta}$-based observation ${\mathbf{y}}_{{n_z}}^\zeta $ can be recast to
\begin{equation}\label{varphi_CS}
{\mathbf{y}}_{{n_z}}^\zeta  = {\mathbf{B}}_{{n_z}}^\zeta {\mathbf{\Psi }}_{{n_z}}^\zeta \left( {\Delta \bm{\zeta }} \right){\mathbf{u}}_{{n_z}}^\zeta  + {\mathbf{B}}_{{n_z}}^\zeta {\mathbf{n}}_{{n_z}}^\zeta .
\end{equation}

\subsection{CS Formulation of $\bf{r}$}
An arbitrary vector indexed by $n_z$ can be drawn out from the $\bf{r}$-based covariance matrix ${{\mathbf{W}}^r} $, i.e., $ {\mathbf{w}}_{{n_y}}^r = {\left[ {{{\left( {{{\mathbf{W}}^r}} \right)}^T}} \right]_{:,{n_y}}} \in {\mathbb{C}^{{N_z} \times 1}} $. To facilitate the gridding over distance, we likewise introduce a uniform grid of   discrete points of $\bf{r}$, i.e., 
\begin{equation}\label{r_grid}
	\resizebox{\hsize}{!}{$
	\left\{ {{{\tilde r}_{{n_z}}}:{{\tilde r}_{{n_z}}} = \frac{{{Z_\Delta }}}{{{n_z} + 1}}\left( {1 - {{\left( {g_l^{\left( {{n_y},{n_z}} \right)}} \right)}^2}} \right),{n_z} = 0,...,{{\tilde N}_z} - 1} \right\}$},
\end{equation}
where $ {Z_\Delta } = \frac{{N_y^2}}{{8\beta _\Delta ^2\delta }}\mathop  \to \limits^{\delta  = \frac{\lambda }{4}} \frac{{N_y^2}}{{2\beta _\Delta ^2\lambda }} $ is the minimum value to ensure the approximate orthogonality of the dictionary matrix with  $ {\beta _\Delta } $ representing the orthogonality controlling factor in order to make sure the grid coherence being below $ \frac{1}{2} $ \cite{XLM-1}. Taking into account the upper and lower bounds of the Rayleigh distance, i.e., $ \left[ {0.5\sqrt {\frac{{{D^3}}}{\lambda }} ,\frac{{2{D^2}}}{\lambda }} \right] $ ($D$ is the array aperture), the number of the grid points can be determined in accordance with $ {\tilde r_{{n_z}}} = \frac{{{Z_\Delta }}}{{{n_z} + 1}}\left( {1 - {{\left( {g_l^{\left( {{n_y},{n_z}} \right)}} \right)}^2}} \right) \ge {\mathfrak{D}_{\min }} $ ($ {\mathfrak{D}_{\min }} = {0.5\sqrt {\frac{{{D^3}}}{\lambda }} } $), and the maximum of $\tilde{N} _z $ takes the value of
\begin{equation}
	 {\left( {{{\tilde N}_z}} \right)_{\max }} = \left\lfloor {\frac{{N_y^2}}{{2\beta _\Delta ^2\lambda {\mathfrak{D}_{\min }}}}\left( {1 - {{\left( {g_l^{\left( {{n_y},{n_z}} \right)}} \right)}^2}} \right)} \right\rfloor .
\end{equation}
Furthermore, the $\bf{r}$-based offset vector  $ \Delta {\mathbf{r}} = \left\{ {\Delta {r_1},...,\Delta {r_{{{\tilde N}_z}}}} \right\} $ is specified with its $\tilde{n} _z$th entry given~by
\begin{equation}
\Delta {r_{{n_z}}} = \left\{ \begin{gathered}
	{r_{{n_z}}} - {{\tilde r}_{{n_z}}}, \text{if} \ {n_z} = {n_{z,q}},q = 1,...L, \hfill \\
	0,\qquad \quad \ {\text{otherwise}}, \hfill \\ 
\end{gathered}  \right.
\end{equation}
where  $ {n_{z,q}} $ indicates the index of the distance grid point nearest to the distance of the $l$th path. Then, denoting $ {\mathbf{y}}_{{n_y}}^r $ and $ {\mathbf{u}}_{{n_y}}^r $ by the $\bf{r}$-based observation vector and the sparse vector, respectively, we can likewise structure ${\mathbf{w}}_{{n_y}}^r$ as its CS form

\begin{equation}\label{r_CS}
	{\mathbf{y}}_{{n_y}}^r = {\mathbf{B}}_{{n_y}}^r{\mathbf{\Psi }}_{{n_y}}^r\left( {\Delta {\mathbf{r}}} \right){\mathbf{u}}_{{n_y}}^r + {\mathbf{B}}_{{n_y}}^r{\mathbf{n}}_{{n_y}}^r,
\end{equation}
where $ {\mathbf{B}}_{{n_y}}^r \in {\mathbb{C}^{{N_{{\text{RF}}}} \times {N_z}}} $ is the size-reducing measurement matrix,  $ {\mathbf{n}}_{{n_y}}^r $ is the noise vector, and  $ {\mathbf{\Psi }}_{{n_y}}^r\left( {\Delta {\mathbf{r}}} \right) \in {\mathbb{C}^{{N_z} \times {{\tilde N}_z}}} $, $ {\mathbf{\Psi }}_{{n_y}}^r\left( {\Delta {\mathbf{r}}} \right) \in {\mathbb{C}^{{N_z} \times {{\tilde N}_z}}} $ is the dictionary matrix, whose  $\left( {n_z, \tilde n_z}  \right) $th entry is given by
\begin{align}
&{\left[ {{\mathbf{\Psi }}_l^r\left( {\Delta {{\mathbf{r}}_{{n_y}}}} \right)} \right]_{{n_z},{{\tilde n}_z}}} \nonumber\\
&= \exp \left( {\jmath \pi \left[ {\frac{{n_y^2}}{{{{\tilde r}_{{n_z}}} + \Delta {r_{{n_z}}}}} - \frac{1}{2}{n_y}\sin {\theta _l}\cos {\varphi _l}} \right]} \right). 
\end{align}

\subsection{Discussion}

Our goal is to obtain the offset vectors $\Delta \bm{\theta}$, $\Delta \bm{\varphi}$ and $\Delta \bf{r}$ from the  3D AED parameter-based CS problem given in (\ref{theta_CS_2}), (\ref{varphi_CS}), and (\ref{r_CS}), respectively. Then, the 3D AED parameter can be recovered by $ {\hat \theta _l} = {\tilde \theta _{{n_{z,q}}}} + \Delta {\theta _{{n_z}}} $, $ {\hat \varphi _l} = \arcsin \frac{{\sin \left( {{{\tilde \zeta }_{{n_{y,q}}}} + \Delta {\zeta _{{n_y}}}} \right)}}{{\sin {{\hat \theta }_l}}} $, and $ {\hat r_l} = {\tilde r_{{n_{z,q}}}} + \Delta {r_{{n_z}}},q = 1,...,L $. The way to accurately acquire the offset vectors will be elaborated in the next section. Let us look first at a pair of two issues, in which how the proposed DeRe paradigm is a curse when concerned in terms of angular index  misinterpretations but a blessing when embraced with regard to significantly reduced gridding complexity.

\subsubsection{Angular Index Correction}

\begin{figure}
	\subfigure[]{
		\begin{minipage}[t]{0.47\textwidth}
			\centering
			\includegraphics[width=\textwidth]{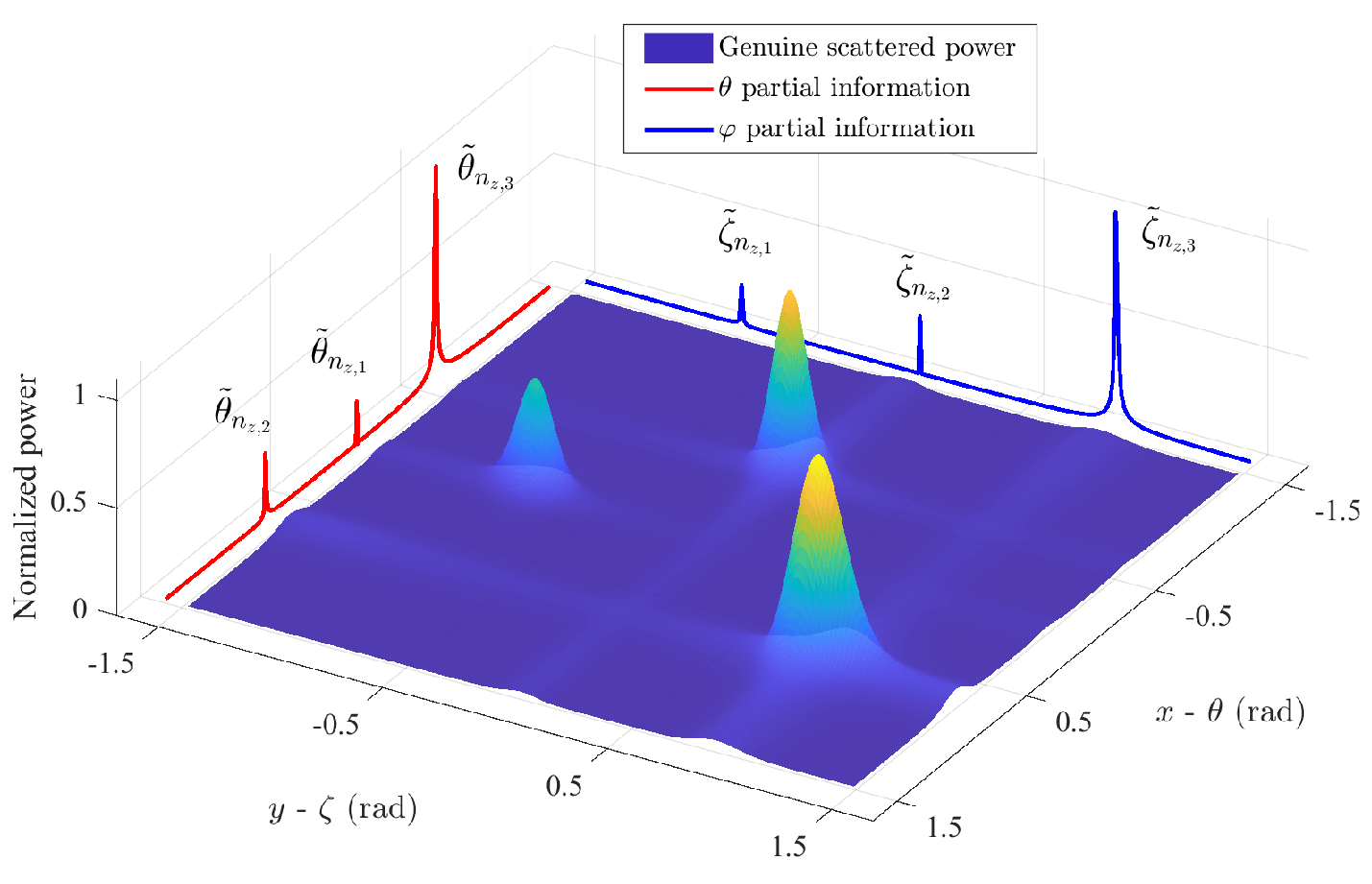}
			\label{AIC1}
	\end{minipage}}
	\subfigure[]{
		\begin{minipage}[t]{0.44\textwidth}
			\centering
			\includegraphics[width=\textwidth]{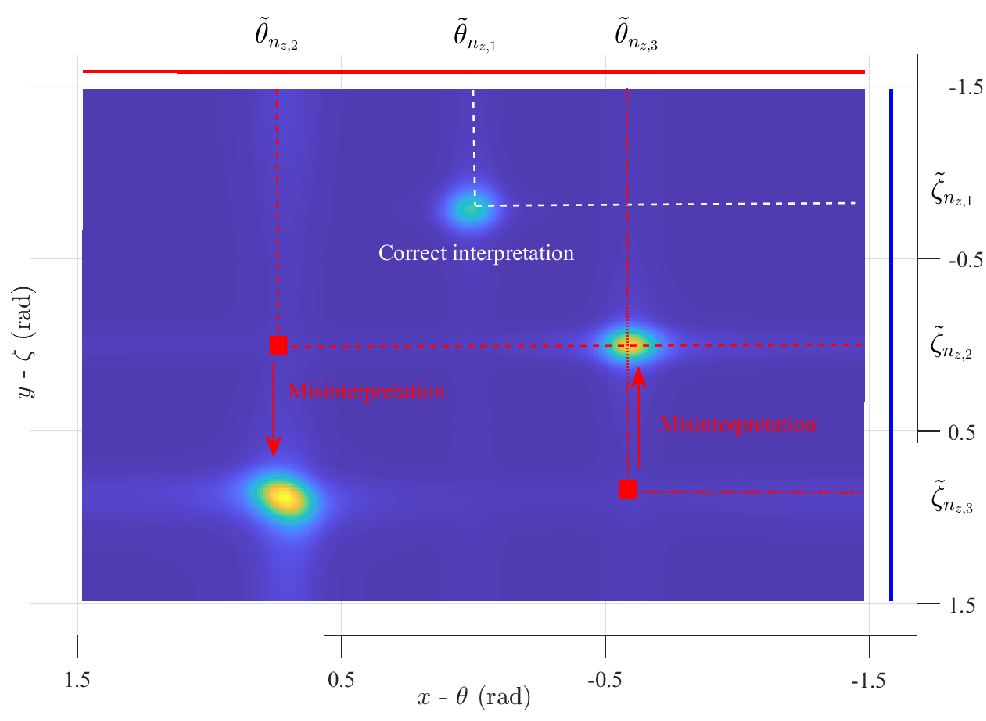}
			\label{AIC2}
	\end{minipage}}
	\caption{(a) Normalized power of the azimuth-elevation angle pair $\left( \bm{\theta}, \bm{\varphi} \right) $, and their independent partial information projected on the $ xoz $ and $ yoz $ plane. (b) Misinterpretations between the attained angular indices and the true indices of the significant angles.}\label{AIC}
\end{figure}

The outputs of the DeRe framework, i.e., $\bm{ \theta }$, $\bm{\varphi}$, and $\bf{r}$, just play separate roles on a significant path, which can be referred to as the partial information. When recovering the original holographic channel using the obtained independent partial information, there typically exist the angular index misinterpretations which significantly deteriorate the estimation performance. To elaborate, in examination of case $L=3$, Fig.~\ref{AIC1} portrays the genuine angular power scattered in relation to the azimuth-elevation angle pair $\left( \bm{\theta}, \bm{\varphi} \right) $ after DeRe filtering, as well as the misinterpretation between the attained angular indices and the true indices of the significant angles when used with partial information. There exist three zero-entries in each of the sparse vectors $ {\left[ {{\mathbf{u}}_{{n_y}}^\theta } \right]_{{n_{z,q}}}} $ and $ 	{\left[ {{\mathbf{u}}_{{n_z}}^\zeta } \right]_{{n_{y,q}}}} $. By retrieving the indices of the non-zero entries for this pair of  sparse vectors, the true index of the $l$th significant angle pair $ \left( \theta_l , \varphi_l\right)  $ may be adversely misinterpreted for that of other significant angle pairs. For example, having retrieved the index $q=2$ for $ \tilde{\theta} _ {n_z,2} $ and $ \tilde{\zeta} _ {n_y,2} $, we may obtain a significant angle pair $ \tilde{\theta} _ {n_z,2} $ and $ \tilde{\zeta} _ {n_y,2} $ (with red  marker $ \square $ in Fig.~\ref{AIC2}). However, this results in a false index for the genuine significant angle pair associated with the path whose true index is $l=3$. Such a misinterpretation significantly erodes the estimation accuracy. To address this hiccup, it is crucial to conduct the angular index correction after attaining the estimates of $\tilde \theta_l$  and $\tilde \zeta_l$.

Regarding the $n_z$th entry in the $\bm{ \theta }$-based sparse vector $ {\mathbf{u}}_{{n_y}}^\theta  $, we denote $q$ by its index of the non-zero entry, whose $q$th non-zero entry is given by
\begin{equation}\label{u_n_y_q}
{\left[ {{\mathbf{u}}_{{n_y}}^\theta } \right]_{{n_{z,q}}}} = 2\bar \beta _l^2\cos \left( {\pi {n_y}\sin {\theta _l}\sin {\varphi _l}} \right),q = 1,...,L.
\end{equation}
Similarly, the non-zero entry in the $\bm{\varphi}$-based sparse vector $ {{\mathbf{u}}_{{n_z}}^\zeta } $ indexed by $q$ is given by
\begin{equation}\label{u_n_z_q}
{\left[ {{\mathbf{u}}_{{n_z}}^\zeta } \right]_{{n_{y,q}}}} = 2\bar \beta _l^2\cos \left( {\pi {n_z}\cos {\theta _l}} \right),q = 1,...,L.
\end{equation}
Since there are $L$ paths in total, there will be $L!$ misinterpretation combinations. We are trying to identify the true indices of the azimuth and elevation angles associated with the $ l $th path. To this end, a pair of variables $ {{\mathfrak{e}^\theta }} $ and $ {{\mathfrak{e}^\zeta }} $ are introduced to represent the power gain of the $l$th path (i.e., $ \bar \beta _l^2 $) associated with the decomposed $\bm{\theta}$ and $\bm{\zeta}$, respectively.
Given $ {\hat \theta _l} = {\tilde \theta _{{n_{z,q}}}} + \Delta {\theta _{{n_z}}} $ and $ {\hat \varphi _l} = \arcsin \frac{{\sin \left( {{{\tilde \zeta }_{{n_{y,q}}}} + \Delta {\zeta _{{n_y}}}} \right)}}{{\sin {{\hat \theta }_l}}},q = 1,...,L $, we have
\begin{align}
{\left( {{\mathfrak{e}^\theta }} \right)_{{n_{z,q}}}} 
&= \frac{1}{{{N_y}}}\sum\limits_{{n_y} = 1}^{{N_y}} {\frac{{{{\left[ {{\mathbf{u}}_{{n_y}}^\theta } \right]}_{{n_{z,q}}}}}}{{2\cos \left( {\pi {n_y}\sin {{\hat \theta }_l}\sin {{\hat \varphi }_l}} \right)}}} \nonumber\\
& \resizebox{0.85\hsize}{!}{$
= \frac{1}{{{N_y}}}\sum\limits_{{n_y} = 1}^{{N_y}} {\frac{{{{\left[ {{\mathbf{u}}_{{n_y}}^\theta } \right]}_{{n_{z,q}}}}}}{{2\cos \left( {\pi {n_y}\sin \left( {{{\tilde \zeta }_{{n_{y,q}}}} + \Delta {\zeta _{{n_y}}}} \right)} \right)}}} ,q = 1,...,L, $} 
\end{align}
\begin{align}
	\resizebox{\hsize}{!}{$
{\left( {{\mathfrak{e}^\zeta }} \right)_{{n_{y,q}}}} = \frac{1}{{{N_z}}}\sum\limits_{{n_z} = 1}^{{N_z}} {\frac{{{{\left[ {{\mathbf{u}}_{{n_z}}^\zeta } \right]}_{{n_{y,q}}}}}}{{2\cos \left( {\pi {n_z}\cos \left( {{{\tilde \theta }_{{n_{z,q}}}} + \Delta {\theta _{{n_z}}}} \right)} \right)}}} ,q = 1,...,L. $}
\end{align}
The idea of the angular index correction is straightforward, that is, identifying the two entries with the closest power, and matching their indices to the index of the path. Therefore, we can accurately retrieve the exact index of the $l$th path associated with the azimuth-elevation angle pair ($\theta_l, \varphi_l$) 
\begin{equation}\label{correction}
\hat q = \arg \min {\left\| {{{\left( {\mathfrak{e}^\theta } \right)}_{{n_{z,q}}}} - {{\left( {\mathfrak{e}^\zeta } \right)}_{{n_{y,q}}}}} \right\|^2}.
\end{equation}
Having identified the true index $\hat{q}$, the power gain $ \bar \beta _l^2 $ can be obtained by $ \bar \beta _l^2 = {\left( {{\mathfrak{e}^\theta }} \right)_{{n_{z,\hat q}}}} = {\left( {{\mathfrak{e}^\zeta }} \right)_{{n_{y,\hat q}}}},\hat q = 1,...,L $.

\subsubsection{Gridding Complexity}
The use of the DeRe framework is capable of significantly reducing the gridding complexity. In particular, gridding across the coupled 3D AED parameters has a complexity order of $\mathcal{O} \left( {N^3} \right) $, in conjunction with $N = N_y \times N_z$. Consequently, the estimation process would be less efficient owing to the resultant extremely large search space. By contrast, 1D gridding over $\bm{ \theta }$ has a complexity order of $\mathcal{O} \left( {N_z}\right) $, and gridding over both $\bm{ \varphi }$ and $\bf{r}$ shares the same complexity order of $\mathcal{O} \left( {N_y}\right) $, which leads to a total gridding complexity order of $\mathcal{O} \left( {N_z + 2N_y} \right) $. Therefore, it is intuitive that the original multiplicative gridding complexity is relaxed to its additive counterpart, significantly lowering the computational burden in practice.

\subsection{Robustness of DeRe Framework}

Given the potential adverse effects of hardware imperfections, such as deviations between the practical and theoretical designs of the combining matrix $ \bf{F} $ and holographic pattern $ \bf{M} $, this subsection aims for examining the robustness of the DeRe framework against such perturbations in the context of channel estimation.
Note that ${\bf{y}}^\theta _{n_y}$ is actually a fictitious vector for ease of analysis (as are ${\bf{w}}^\theta _{n_y}$ and $\bf{V}$), so before proceeding to unveiling DeRe's robustness, let us establish a critical bridge between the physical signal received from the BS and the observation ${\bf{y}}^\theta _{n_y}$ utilized in the CS procedure. Subsequently, we shall take $\bm{ \theta }$ as an illustrative instance.

Let $ {{\mathbf{y}}^\tau },\tau  \in \left\{ {1,...,T} \right\} $ denote an arbitrary sample of the received signal at the BS. We define $ {f_{a \to b}}\left( a \right) $ as the mapping from $a$ to $b$. For an arbitrary observation sampled at $\tau$, we obtain $ {\mathbf{y}}_{{n_y}}^{\theta ,\tau } = {\mathbf{B}}_{{n_y}}^{\theta ,\tau }{\mathbf{w}}_{{n_y}}^{\theta ,\tau } $, whose mapping relation is given by $ {f_{{\mathbf{w}} \to {\mathbf{y}}}}\left( {{\mathbf{w}}_{{n_y}}^{\theta ,\tau }} \right) = {f_{{\mathbf{w}} \to {\mathbf{y}}}}\left( {{f_{{\mathbf{V}} \to {\mathbf{w}}}}\left( {\mathbf{V}} \right)} \right) = {f_{{\mathbf{w}} \to {\mathbf{y}}}}\left( {{f_{{\mathbf{V}} \to {\mathbf{w}}}}\left( {{f_{{\mathbf{h}} \to {\mathbf{V}}}}\left( {{{\mathbf{h}}^\tau }} \right)} \right)} \right) $. The mapping $ {f_{{\mathbf{w}} \to {\mathbf{y}}}}\left( {{f_{{\mathbf{V}} \to {\mathbf{w}}}}\left( {\mathbf{V}}^\tau \right)} \right) $ can be recast by means of a matrix manipulation given by
	\begin{equation}
		{f_{{\mathbf{V}} \to {\mathbf{w}}}}\left( {{{\mathbf{V}}^\tau }} \right) = {{\mathbf{C}}_{{\mathbf{V}} \to {\mathbf{w}}}}\left( {{{\mathbf{I}}_N} + {{\mathbf{E}}_{{\mathbf{V}} \to {\mathbf{w}}}}} \right){{\mathbf{V}}^\tau },
	\end{equation}
	where $ {{\mathbf{E}}_{{\mathbf{V}} \to {\mathbf{w}}}} = \left[ {{\mathbf{\tilde e}}_1^T;...;{\mathbf{\tilde e}}_N^T} \right] \in {\mathbb{C}^{N \times N}} $, with the $n$th vector $ {{\mathbf{\tilde e}}_n} $ given by ${{\mathbf{\tilde e}}_n} = {{\mathbf{e}}_{n^\prime}}$ and  $ {{\mathbf{e}}_{n'}} \in {\mathbb{R}^{N \times 1}} $ representing the basis vector whose $n^\prime$th entry is non-zero. The matrix $ {{\mathbf{C}}_{{\mathbf{V}} \to {\mathbf{w}}}} \in {\mathbb{C}^{N_z \times N}}$ is constituted by $ \left[ {{{\mathbf{c}}_1},...,{{\mathbf{c}}_N}} \right]  $, whose $n$th vector is crafted by
	\begin{equation}
		{{\mathbf{c}}_n} \in {\mathbb{R}^{{N_z} \times 1}} = \left\{ \begin{gathered}
			{{\mathbf{e}}_i},n = \left( {{n_y},i} \right),i = 1,...,{N_z}, \hfill \\
			{\mathbf{0}}, \ {\text{others.}}\hfill \\ 
		\end{gathered}  \right.
	\end{equation}
	It thus holds that $ {f_{{\mathbf{w}} \to {\mathbf{y}}}}\left( {{f_{{\mathbf{V}} \to {\mathbf{w}}}}\left( {{f_{{\mathbf{h}} \to {\mathbf{V}}}}\left( {{{\mathbf{h}}^\tau }} \right)} \right)} \right) = {\mathbf{B}}_{{n_y}}^{\theta ,\tau }{{\mathbf{C}}_{{\mathbf{V}} \to {{\mathbf{w}}^\theta }}}\left( {{{\mathbf{I}}_N} + {{\mathbf{E}}_{{\mathbf{V}} \to {{\mathbf{w}}^\theta }}}} \right)\left( {\text{diag}} \left( {{{\mathbf{h}}^\tau }} \right) \right){{\mathbf{E}}_{{\mathbf{h}} \to {\mathbf{V}}}}{{\mathbf{h}}^{*,\tau }} $, in which the implementation of ${{\mathbf{E}}_{{\mathbf{h}} \to {\mathbf{V}}}}$ is similar to $ {{\mathbf{E}}_{{\mathbf{V}} \to {\mathbf{w}}}} $'s, just using different index selection strategy shown in (\ref{v_n}). As a result, the received signal at the BS that serves as an observation  $ {\mathbf{y}}_{{n_y}}^\theta  $ can be given by
	\begin{align}\label{mapping_y_theta}
		{\mathbf{y}}_{{n_y}}^\theta  = \mathbb{E}\left\{ {{\mathbf{B}}_{{n_y}}^{\theta ,\tau }{{\mathbf{C}}_{{\mathbf{V}} \to {{\mathbf{w}}^\theta }}}\left( {{{\mathbf{I}}_N} + {{\mathbf{E}}_{{\mathbf{V}} \to {{\mathbf{w}}^\theta }}}} \right){\text{diag}}\left( {{{\mathbf{h}}^\tau }} \right){{\mathbf{E}}_{{\mathbf{h}} \to {\mathbf{V}}}}{{\mathbf{h}}^{*,\tau }}} \right\} \nonumber\\
		= {\mathbf{B}}_{{n_y}}^\theta {{\mathbf{C}}_{{\mathbf{V}} \to {{\mathbf{w}}^\theta }}}\left( {{{\mathbf{I}}_N} + {{\mathbf{E}}_{{\mathbf{V}} \to {{\mathbf{w}}^\theta }}}} \right)\mathbb{E}\left\{ {{\text{diag}}\left( {{{\mathbf{h}}^\tau }} \right){{\mathbf{E}}_{{\mathbf{h}} \to {\mathbf{V}}}}{{\mathbf{h}}^{*,\tau }}} \right\}.
	\end{align}
In practical systems, the linear transformation in (\ref{mapping_y_theta}), i.e., $ {\mathbf{B}}_{{n_y}}^\theta {{\mathbf{C}}_{{\mathbf{V}} \to {{\mathbf{w}}^\theta }}}\left( {{{\mathbf{I}}_N} + {{\mathbf{E}}_{{\mathbf{V}} \to {{\mathbf{w}}^\theta }}}} \right) $ can be achieved by appropriately designing the combining matrix and the holographic pattern. The remnant expectation term $ \mathbb{E}\left\{ {{\text{diag}}\left( {{{\mathbf{h}}^\tau }} \right){{\mathbf{E}}_{{\mathbf{h}} \to {\mathbf{V}}}}{{\mathbf{h}}^{*,\tau }}} \right\} $ can be realized by an output of the Hadamard product whose inputs take the signal received at the BS and its conjugate counterpart.

The structure in (\ref{mapping_y_theta}) indicates that the signal received at the BS can be crafted to attain the observation $ {\mathbf{y}}_{{n_y}}^\theta $ embracing $\bm{\theta}$-associated statistical characteristics. Regarding the error induced by hardware imperfections, we let $ {\mathbf{B}}_{{n_y}}^{\theta ,\tau }{{\mathbf{C}}_{{\mathbf{V}} \to {{\mathbf{w}}^\theta }}}\left( {{{\mathbf{I}}_N} + {{\mathbf{E}}_{{\mathbf{V}} \to {{\mathbf{w}}^\theta }}}} \right) = {\mathbf{\bar F}} + \Delta {\mathbf{\bar F}} $, where $ {\mathbf{\bar F}} $ denote the intended design of the combining matrix and holographic pattern, while $ \Delta {\mathbf{\bar F}} $  captures the deviation between the actual implementation and the intended design. Accordingly, (\ref{mapping_y_theta}) can be recast as 
\begin{align}\label{mapping_y_theta2}
{\mathbf{y}}_{{n_y}}^\theta & = \mathbb{E}\left\{ {\left( {{\mathbf{\bar F}} + \Delta {\mathbf{\bar F}}} \right){\text{diag}}\left( {{{\mathbf{h}}^\tau }} \right){{\mathbf{E}}_{{\mathbf{h}} \to {\mathbf{V}}}}{{\mathbf{h}}^{*,\tau }}} \right\} \nonumber\\
& = \mathbb{E}\left\{ {\left( {{\mathbf{\bar F}} + \Delta {\mathbf{\bar F}}} \right)} \right\}\mathbb{E}\left\{ {{\text{diag}}\left( {{{\mathbf{h}}^\tau }} \right){{\mathbf{E}}_{{\mathbf{h}} \to {\mathbf{V}}}}{{\mathbf{h}}^{*,\tau }}} \right\} \nonumber\\
 & = \left( {{\mathbf{\bar F}} + \mathbb{E}\left\{ {\Delta {\mathbf{\bar F}}} \right\}} \right)\mathbb{E}\left\{ {{\text{diag}}\left( {{{\mathbf{h}}^\tau }} \right){{\mathbf{E}}_{{\mathbf{h}} \to {\mathbf{V}}}}{{\mathbf{h}}^{*,\tau }}} \right\}.
\end{align}
The form in (\ref{mapping_y_theta2}) is appealing since it is analytically tractable. This implies that the resultant error can be mitigated through skillful designs of the combining matrix and holographic pattern. Specific design details are beyond the scope of this paper, and interested readers may refer to \cite{HW-4,XLM-7} for inspiration in transceiver design within the context of holographic scope. The robustness performance of the DeRe framework will be evaluated later in Sec.~VI.

\section{DeRe-VM Algorithm}

In pursuit of an accurate and robust channel estimation for holographic MIMO systems, we intend to recover the sparse vectors ${\bf{u}}^\theta_{n_y}$, ${\bf{u}}^\zeta_{n_y}$ and ${\bf{u}}^r_{n_y}$, as well as their associated offset vectors $\Delta \bm{\theta}$, $\Delta \bm{\zeta}$ and $\Delta \bf{r}$, as presented in (\ref{theta_CS_2}), (\ref{varphi_CS}), and (\ref{r_CS}), respectively. Albeit the seeming straightforwardness of that kind of parameter-recovery problem, it poses arduous challenges due to a pair of peculiarities: i) the uncertainty of the sensing matrices including the offsets to be determined, i.e., $ {\mathbf{\Psi }}_{{n_y}}^\theta \left( {\Delta \bm{\theta}} \right) $, $  {\mathbf{\Psi }}_{{n_y}}^\zeta \left( {\Delta \bm{\zeta}} \right)$, and $ {\mathbf{\Psi }}_{{n_y}}^r \left( {\Delta \bf{r}} \right)$, as well as ii) the imperfect priors of the sparse vectors induced by the unavailability of the exact distributions for  ${\bf{u}}^\theta_{n_y}$, ${\bf{u}}^\zeta_{n_y}$, and ${\bf{u}}^r_{n_y}$. Therefore, targeting the twofold issues, a layered probability model is leveraged to provide appropriate prior distributions for the sparse vectors, followed by an efficient and robust algorithm design.

\subsection{Layered Probability Model}

For the sake of brevity, we take $ \bm{\theta}  $ as an example to explicate the way to harness the structured sparsity underlying  ${\bf{w}}^\theta _ {n_y}$.
Concerning that $ {N_y} $  estimates of  $ \bm{\theta} $ can be attained by resolving  $N_y$ copies of problem (\ref{theta_CS_2}), we can group the sparse vectors $ {\mathbf{u}}_1^\theta ,...,{\mathbf{u}}_{{n_y}}^\theta ,...,{\mathbf{u}}_{{N_y}}^\theta  $  into $ {\tilde N_z} $  blocks with block size $ {N_y} $, in which the  $n_z$th block is given by  $ \left[ {u_{1,{n_z}}^\theta ,...,u_{{n_y},{n_z}}^\theta ,...,u_{{N_y},{n_z}}^\theta } \right] $. Let $\bm{\rho }_{}^\theta  = {\left[ {\bm{\rho }_1^\theta ,...,\bm{\rho }_{{N_y}}^\theta } \right]^T}$ represent the precision vector, in which $1/\rho _{{n_y},{n_z}}^\theta $ denotes the variance of $u_{{n_y},{n_z}}^\theta $. Furthermore, a support vector denoted by ${\bm{\alpha }^\theta } \in {\left\{ {0,1} \right\}^{{{\tilde N}_z}}}$  is introduced, whose  $n_z$th entry $\alpha _{{n_z}}^\theta $ indicates whether the $n_z$th block is active ($\alpha _{{n_z}}^\theta  = 1$) or not ($\alpha _{{n_z}}^\theta  = 0$). Therefore, the joint distribution of $ {{\mathbf{u}}^\theta } $, $ \bm{\rho }_{}^\theta  $, and ${\bm{\alpha }^\theta }$ is in the form of
\begin{equation}\label{joint_distribution}
p\left( {{\mathbf{u}}_{{n_y}}^\theta ,\bm{\rho }_{{n_y}}^\theta ,{\bm{\alpha }^\theta }} \right) = \underbrace {p\left( {{\bm{\alpha }^\theta }} \right)}_{{\text{Support}}}\underbrace {p\left( {\bm{\rho }_{{n_y}}^\theta |{\bm{\alpha }^\theta }} \right)}_{{\text{Precision}}}\underbrace {p\left( {{\mathbf{u}}_{{n_y}}^\theta \bm{|\rho }_{{n_y}}^\theta } \right)}_{{\text{Sparse vector}}},
\end{equation}
Then let us move on to describe in greater detail the prior distribution of each layer.

\subsubsection{Layer I: Probability Model for Support Vector $ {\bm{\alpha }^\theta } $}

Due to the limited scatterers in the physical propagation environment, the birth-death process of ${\bm{\alpha }^\theta }$ can be modeled as a Markov prior
\begin{equation}
p\left( {{\bm{\alpha }^\theta }} \right) = p\left( {\alpha _{{n_z}}^\theta } \right)\prod\limits_{{n_z} = 2}^{{{\tilde N}_z}} {p\left( {\alpha _{{n_z}}^\theta |\alpha _{{n_z} - 1}^\theta } \right)} ,
\end{equation}
with the transition probability $p\left( {\alpha _{{n_z}}^\theta |\alpha _{{n_z} - 1}^\theta } \right)$ given by 
\begin{equation}\label{Markov}
p\left( {\alpha _{{n_z}}^\theta |\alpha _{{n_z} - 1}^\theta } \right) = \left\{ \begin{gathered}
	{\left( {1 - {p_{0 \to 1}}} \right)^{1 - \alpha _{{n_z}}^\theta }}{\left( {{p_{0 \to 1}}} \right)^{\alpha _{{n_z}}^\theta }},\alpha _{{n_z} - 1}^\theta  = 0 ,\hfill \\
	{\left( {{p_{1 \to 0}}} \right)^{1 - \alpha _{{n_z}}^\theta }}{\left( {1 - {p_{1 \to 0}}} \right)^{\alpha _{{n_z}}^\theta }},\alpha _{{n_z} - 1}^\theta  = 1, \hfill \\ 
\end{gathered}  \right.
\end{equation}
where ${p_{0 \to 1}}$ represents the transition probability of entries in the sparse vector ${\mathbf{u}}_{{n_y}}^\theta $ shifting from 0 to 1, and vice versa for ${p_{0 \to 1}}$.

\subsubsection{Layer II: Probability Model for Precision Vector ${\bm{\rho }^\theta }$}
A Gamma prior distribution is imposed on the precision vector ${\bm{\rho }^\theta }$ of the support vector ${\bm{\alpha }^\theta }$
\begin{align}
p\left( {{\bm{\rho }^\theta }|{\bm{\alpha }^\theta }} \right) &= \underbrace {\prod\limits_{{n_z} = 1}^{{{\tilde N}_z}} {p\left( {\bm{\rho }_{{n_z}}^\theta |\alpha _{{n_z}}^\theta } \right)} }_{f_{{n_z}}^{\bm{\rho }}} = \prod\limits_{{n_y} = 1}^{{N_y}} {\prod\limits_{{n_z} = 1}^{{{\tilde N}_z}} {p\left( {\rho _{{n_y},{n_z}}^\theta |\alpha _{{n_z}}^\theta } \right)} }  \nonumber\\
 &= \prod\limits_{{n_y} = 1}^{{N_y}} \prod\limits_{{n_z} = 1}^{{{\tilde N}_z}} {{\left[ {\Gamma \left( {\rho _{{n_y},{n_z}}^\theta ;\mathfrak{a}_{{n_y},{n_z}}^\theta ,\mathfrak{b}_{{n_y},{n_z}}^\theta } \right)} \right]}^{\alpha _{{n_z}}^\theta }} \nonumber\\
 &\quad \times {{\left[ {\Gamma \left( {\rho _{{n_y},{n_z}}^\theta ;\bar {\mathfrak{a}}_{{n_y},{n_z}}^\theta ,\bar {\mathfrak{b}}_{{n_y},{n_z}}^\theta } \right)} \right]}^{1 - \alpha _{{n_z}}^\theta }}  ,
\end{align}
where $ \Gamma \left( {\rho; \mathfrak{a}, \mathfrak{b}} \right)  $ is a Gamma hyperprior with the shape parameter $\mathfrak{a}$ and rate parameter $\mathfrak{b}$. For the case of $ \alpha _{n_z}^\theta  = 1$, the shape and rate parameters $ {\mathfrak{a}_{{n_y},{n_z}}^\theta } $ and $ {\mathfrak{b}_{{n_y},{n_z}}^\theta } $ are selected to satisfy $ \frac{{\mathfrak{a}_{{n_y},{n_z}}^\theta }}{{\mathfrak{b}_{{n_y},{n_z}}^\theta }} = \mathbb{E}\left[ {\rho _{{n_y},{n_z}}^\theta } \right] $, since the variance $ 1/\rho _{{n_y},{n_z}}^\theta  $ of $  {u_{{n_y},{n_z}}^\theta }$ is equivalent to $ \mathbb{E}\left[ {\rho _{{n_y},{n_z}}^\theta } \right] $ and $ {u_{{n_y},{n_z}}^\theta } $ deviates from zero with a high probability when it is active. On the contrary, we let the shape and rate parameters $ \bar {\mathfrak{a}}_{{n_y},{n_z}}^\theta$ and $\bar {\mathfrak{b}}_{{n_y},{n_z}}^\theta $ be chosen such that $ \frac{{\bar {\mathfrak{a}}_{{n_y},{n_z}}^\theta }}{{\bar {\mathfrak{b}}_{{n_y},{n_z}}^\theta }} = \mathbb{E}\left[ {\rho _{{n_y},{n_z}}^\theta } \right] \gg 1$, conditioned on the event of $ \alpha _{n_z}^\theta  \neq 1$, in which case the variance of $ {u_{{n_y},{n_z}}^\theta } $ tends to be zero when the associated path is inactive.

\subsubsection{Layer III: Probability Model for Sparse Vector ${\bf{u}}^\theta$}

The conditional probability for sparse vector ${\bf{u}}^\theta$ is considered to follow a non-stationary Gaussian prior distribution with distinct precision~$ {\bm{\rho }^\theta } $
\begin{align}
p\left( {{{\mathbf{u}}^\theta }|{\bm{\rho }^\theta }} \right) &= \underbrace {\prod\limits_{{n_y} = 1}^{{N_y}} {\prod\limits_{{n_z} = 1}^{{{\tilde N}_z}} {p\left( {u_{{n_y},{n_z}}^\theta |\rho _{{n_y},{n_z}}^\theta } \right)} } }_{f_{{n_y},{n_z}}^{\mathbf{u}}} \nonumber\\
&= \prod\limits_{{n_y} = 1}^{{N_y}} {\prod\limits_{{n_z} = 1}^{{{\tilde N}_z}} {\mathcal{C}\mathcal{N}\left( {u_{{n_y},{n_z}}^\theta ;0,\left( {1/\rho _{{n_y},{n_z}}^\theta } \right)} \right)} } ,
\end{align}
Additionally, in the presence of the complex Gaussian noise, we have 
\begin{equation}
p\left( {{\mathbf{y}}_{{n_y}}^\theta |{\mathbf{u}}_{{n_y}}^\theta } \right) = \mathcal{C}\mathcal{N}\left( {{\mathbf{y}}_{{n_y}}^\theta ;{\mathbf{B}}_{{n_y}}^\theta {\mathbf{\Psi }}_{{n_y}}^\theta {\mathbf{u}}_{{n_y}}^\theta ,{\kappa ^\theta }{{\mathbf{I}}_{{N_{{\text{RF}}}}}}} \right),
\end{equation}
where ${\kappa ^\theta } = \sigma ^{-2}$ represents the noise precision that follows a Gamma hyperprior $p \left( {\kappa ^\theta }\right)= \Gamma\left( {\kappa ^\theta; {\mathfrak{a}}_\kappa, {\mathfrak{b}}_\kappa}\right) $. The tunable $ {\mathfrak{a}}_\kappa $ and $ {\mathfrak{b}}_\kappa $ tend to zero, i.e., ${\mathfrak{a}}_\kappa, {\mathfrak{b}}_\kappa \to 0$, for approximating a more general hyperprior.

\subsubsection{Channel Recovery Problem Formulation}

Given observations ${{\mathbf{y}}^\theta } = \left[ {{\mathbf{y}}_1^\theta ;...;{\mathbf{y}}_{{N_y}}^\theta } \right] \in {\mathbb{C}^{{N_y}{N_z} \times 1}}$, our objective is to estimate the sparse vector ${\mathbf{u}} = \left[ {{\mathbf{u}}_1^\theta ;...;{\mathbf{u}}_{{N_y}}^\theta } \right] \in {\mathbb{C}^{{N_y}{N_z} \times 1}}$ and the grid offsets $\Delta \bm{\theta }$. More precisely, we intend to compute the MMSE estimates of $u_{{n_y},{n_z}}^\theta $, i.e., $\hat u_{{n_y},{n_z}}^\theta  = \mathbb{E}\left[ {u_{{n_y},{n_z}}^\theta |{{\mathbf{y}}^\theta };\Delta \bm{\theta }} \right]$, with the expectation taken over the conditional marginal posterior
\begin{align}\label{conditional_posterior}
&p\left( {u_{{n_y},{n_z}}^\theta |{{\mathbf{y}}^\theta };\Delta \bm{\theta }} \right) \nonumber\\ & \propto \sum\limits_{{\bm{\alpha }^\theta }} {\int_{ - u_{{n_y},{n_z}}^\theta } {p\left( {{{\mathbf{y}}^\theta },{{\mathbf{u}}^\theta },{\bm{\rho }^\theta },{\bm{\alpha }^\theta };\Delta \bm{\theta }} \right)} } \nonumber\\
 &= \sum\limits_{{\bm{\alpha }^\theta }} {\int_{ - u_{{n_y},{n_z}}^\theta } {p\left( {{{\mathbf{u}}^\theta },{\bm{\rho }^\theta },{\bm{\alpha }^\theta };\Delta \bm{\theta }} \right)p\left( {{{\mathbf{y}}^\theta }|{{\mathbf{u}}^\theta };\Delta \bm{\theta }} \right)} } ,
\end{align}
where $ - u_{{n_y},{n_z}}^\theta $ represents the entry removal with respect to $u_{{n_y},{n_z}}^\theta $ over the sparse vector $\bf{u}$ and $\propto$ refers to the equality after scaling. Furthermore, the uncertain grid offset $\Delta \bm{\hat \theta }$ can be attained by MLE as
\begin{align}\label{MLE}
\Delta \bm{\hat \theta } = \mathop {\arg \max }\limits_{\Delta \bm{\theta }} \ln p\left( {{{\mathbf{y}}^\theta };\Delta \bm{\theta }} \right) 
 = \mathop {\arg \max }\limits_{\Delta \bm{\theta }} \int_{{\bm{v}^\theta }} {\ln p\left( {{{\mathbf{y}}^\theta },{\bm{v}^\theta };\Delta \bm{\theta }} \right)} ,
\end{align}
in which ${\bm{v}^\theta } = \left\{ {{{\mathbf{u}}^\theta },{\bm{\rho }^\theta },{\bm{\alpha }^\theta }} \right\}$ represents the collection of variables associated with the probability model. We thus attain the MMSE estimate of $u_{{n_y},{n_z}}^\theta $ by figuring out the MLE of $\Delta \bm{\hat \theta }$ and the conditional marginal posterior $p\left( {u_{{n_y},{n_z}}^\theta |{{\mathbf{y}}^\theta };\Delta \bm{\theta }} \right)$.

\begin{figure}[t]
	\centering
	\includegraphics[width=0.5\textwidth]{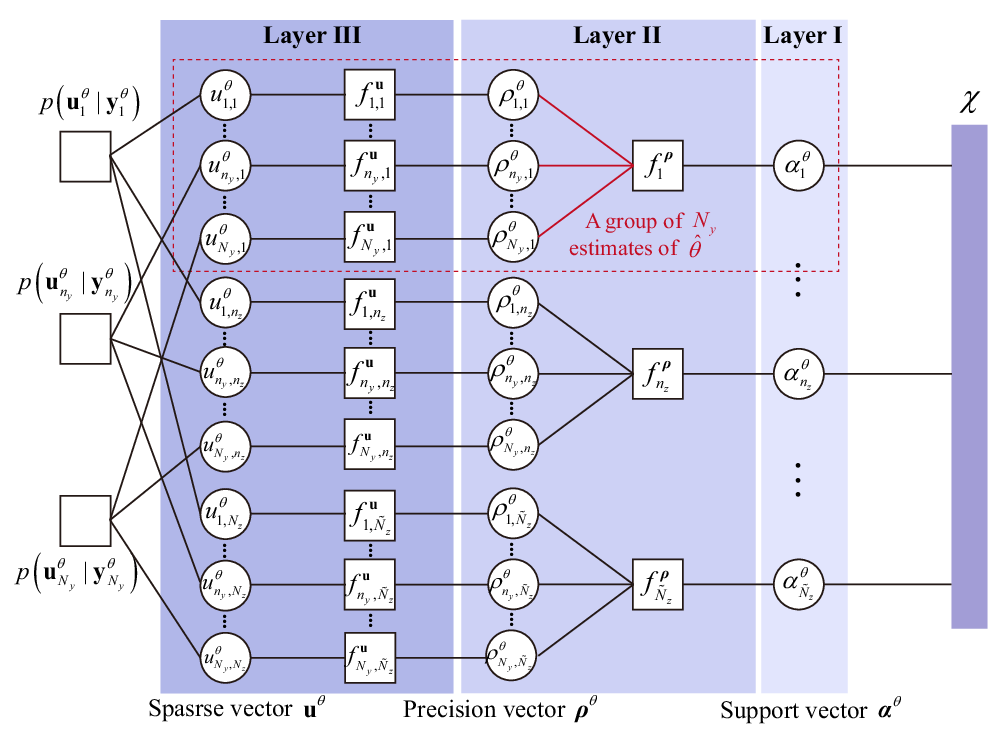}
	\caption{Factor graph of the joint distribution $p\left( {{{\mathbf{y}}^\theta },{\bm{v}^\theta };\Delta \bm{\theta }} \right)$.} \label{FactorGraph}
\end{figure}

Recall that $N_y$ estimates of $\bm{\theta}$ can be attained by solving $N_y$ copies of problem (\ref{theta_CS_2}), and the associated sparse vectors $ {\mathbf{u}}_1^\theta ,...,{\mathbf{u}}_{{n_y}}^\theta ,...,{\mathbf{u}}_{{N_y}}^\theta  $ have been grouped into $ {\tilde N_z} $  blocks with block size $ {N_y} $, as illustrated in the factor graph in Fig.~\ref{FactorGraph}. Such a grouping facilitates an appropriate fusion of $N_y$ beliefs, i.e., messages passing via the path $u_{{n_y},{n_z}}^\theta  \to f_{{n_y},{n_z}}^{\mathbf{u}} \to \rho _{{n_y},{n_z}}^\theta  \to f_{{n_z}}^{\bm{\rho }}, {n_y} = 1,...,{N_y}$.
To be explicit, in light of the belief propagation \cite{Bayesian-15}, the message ${\nu _{f_{{n_z}}^{\bm{\rho }} \to \alpha _{{n_z}}^\theta }}$ is calculated by taking the product of all messages received by the factor node $ f_{{n_z}}^{\bm{\rho }} $ on all other edges, i.e., 
\begin{align}
&{\nu _{f_{{n_z}}^{\bm{\rho }} \to \alpha _{{n_z}}^\theta }}\left( {\alpha _{{n_z}}^\theta } \right) \nonumber\\ &\propto \sum\nolimits_{ - \alpha _{{n_z}}^\theta } {f_{{n_z}}^{\bm{\rho }}\left( {\alpha _{{n_z}}^\theta ,\rho _{1,{n_z}}^\theta , \ldots ,\rho _{{N_y},{n_z}}^\theta } \right)\prod\nolimits_{{n_y} = 1}^{{N_y}} {\rho _{{n_y},{n_z}}^\theta } } .
\end{align}
This implies that each block $n_z$ contains partial priors of the $N_y$ sparse vectors, and priors of $\tilde{N}_z$ blocks function collaboratively on the channel support vector $\bm{\alpha}^{\bm{ \theta }}$.

It is still arduous to obtain the precise estimates of $\hat u_{{n_y},{n_z}}^\theta $ and  $\Delta \bm{\hat \theta }$, since the conditional marginal posterior in (\ref{conditional_posterior}) is unavailable due to the complex integration, and the dense loops in the factor graph exacerbate matters. Additionally, the non-concave nature of problem (\ref{MLE}) hinders the pursuit of its stationary solution even if one uses some plain optimization approaches. Therefore, the subsection that follows proceeds to present beneficial techniques for clearing up the above hindrances, as well as an effective algorithm for facilitating the efficient estimation of $\bm{\theta}$.

\subsection{DeRe-VM Algorithm }
As a key feature of DeRe-VM, an approximate variational distribution $\psi \left( {{\bm{v}^\theta }} \right)$ is employed to refine the intractable posterior $p\left( {{\bm{v}^\theta }|{{\mathbf{y}}^\theta };\Delta \bm{\theta }} \right)$ in the DeRe-VM-E Step, while updating the indeterminate grid offset $ \bm{\theta } $  using the gradient-based technique in the DeRe-VM-M Step.

\subsubsection{DeRe-VM-M Step: Inexact Majorization-Minimization (MM)}

The likelihood function \\ $\ln p\left( {{{\mathbf{y}}^\theta };\Delta \bm{\theta }} \right)$ is indeed intractable owing to the absence of closed-form expressions induced by the prohibitive integrals of the joint distribution with regard to indeterminate ${\bm{v}^\theta }$. An alternative is to construct its surrogate function based on $\Delta {\bm{\theta }^{\left( r \right)}}$ in the $r$th iteration in place of the objective in problem (\ref{MLE}), providing that the exact posterior $ p\left( {{\bm{v}^\theta }|{{\mathbf{y}}^\theta };\Delta {\bm{\theta }^{\left( r \right)}}} \right) $ has been identified. Unfortunately, the exact posterior $p\left( {{\bm{v}^\theta }|{{\mathbf{y}}^\theta };\Delta {\bm{\theta }^{\left( r \right)}}} \right)$ is also highly unavailable in our examined problem owing to the intricate loops inherent in the factor graph. We thus need to leverage the variational Bayesian inference (VBI) and message passing techniques to attain an alternative $\psi \left( {{\bm{v}^\theta }} \right)$ for a faithful approximation of the conditional posterior  $p\left( {{\bm{v}^\theta }|{{\mathbf{y}}^\theta };\Delta {\bm{\theta }^{\left( r \right)}}} \right)$. The to-be-determined posterior $\psi \left( {{\bm{v}^\theta }} \right)$  has a factorized form, i.e.,  $ \psi \left( {{\bm{v}^\theta }} \right) = \prod\nolimits_{j \in \mathcal{J}} {\psi \left( {\bm{v}_j^\theta } \right)}  $, with $\bm{v}_j^\theta $ representing an individual in the variable collection  ${\bm{v}^\theta }$, and we have  $\mathcal{J} = \left\{ {j|\forall \bm{v}_j^\theta  \in {\bm{v}^\theta }} \right\}$. Then, an equivalent of the original MLE problem in (\ref{MLE}) can be obtained based on the approximate posterior, which is formulated as
\begin{equation}\label{surrogate-function}
	\mathop {\max }\limits_{\psi \left( {{{\bm{v}}^\theta }} \right),\Delta \bm{\theta }} \underbrace {\int {\psi \left( {{{\bm{v}}^\theta }} \right)\ln \frac{{p\left( {{{\mathbf{y}}^\theta },{{\bm{v}}^\theta };\Delta {{\bm{\theta }}^{\left( r \right)}}} \right)}}{{\psi \left( {{{\bm{v}}^\theta }} \right)}}} d{{\bm{v}}^\theta }}_{\mathcal{F}\left( {\Delta \bm{\theta };\Delta {{\bm{\theta }}^{\left( r \right)}}} \right)}.
\end{equation}
In the $(r+1)$th iteration in the DeRe-VM-M Step,  $\Delta {\bm{\theta }^{\left( {r + 1} \right)}}$ can be updated by 
\begin{equation}\label{MM-update1}
\Delta {\bm{\theta }^{\left( {r + 1} \right)}} = \mathop {\arg \max }\limits_{\Delta \bm{\theta }} \mathcal{F}\left( {\Delta \bm{\theta };\Delta {\bm{\theta }^{\left( r \right)}}} \right) .
\end{equation}
The non-concave nature of the objective, nevertheless, impedes the pursuit of the global optimum for problem (\ref{MM-update1}). We thus exploit the following gradient update in an effort to obtain a stationary solution 
\begin{equation}\label{update-delta-theta}
\Delta {\bm{\theta }^{\left( {r + 1} \right)}} = \Delta {\bm{\theta }^{\left( r \right)}} + {\iota ^{\left( r \right)}}\frac{{\partial \mathcal{F}\left( {\Delta \bm{\theta };\Delta {\bm{\theta }^{\left( r \right)}}} \right)}}{{\partial \Delta \bm{\theta }}},
\end{equation}
where  $ {\iota ^{\left( r \right)}} $ is the step size determined by the Armijo rule \cite{chen-twc3}.

\begin{figure*}[t]
	\centering
	\includegraphics[width=0.85\textwidth]{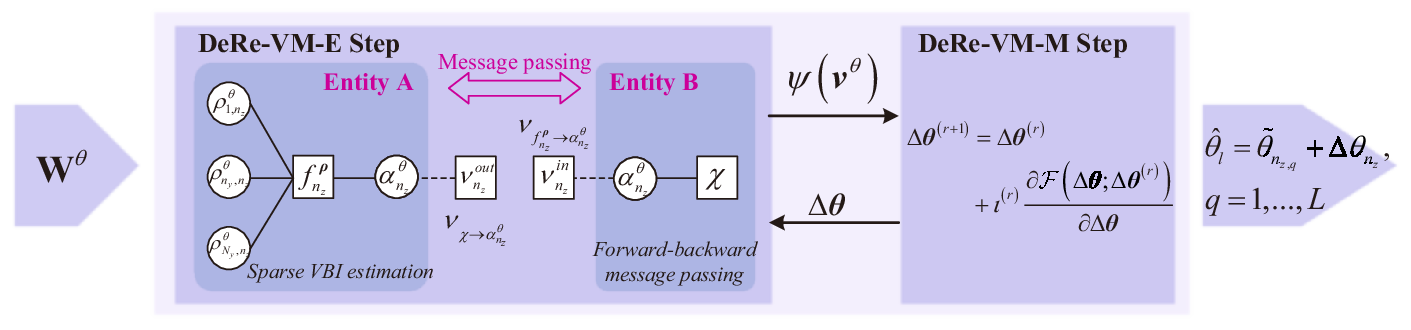}
	\caption{Schematic diagram of the proposed DeRe-VM algorithm by taking $\bm{ \theta }$ as an example.} \label{DeRe-VM}
\end{figure*}

\subsubsection{DeRe-VM-E Step: A Pair of Entities}
Due to myriads of loops in a dense factor graph depicted in Fig.~\ref{FactorGraph}, a direct employment of the classical sum-product message passing (SPMP) \cite{TIT-2001} fails to traverse the factor and variable nodes over the entire factor graph. As potential solutions, the approximate message passing (AMP)-based algorithms, e.g., Turbo-AMP \cite{AngDom-11-11} and Turbo-CS \cite{AngDom-11-12}, easily get entrapped into their poor local optimum due to the ill-conditioned dictionary matrix  ${\mathbf{\Psi }}_{}^\theta \left( {\Delta \bm{\theta }} \right)$ induced by the indeterminate $ {\Delta \bm{\theta }} $. To facilitate the practical implementation, two entities are specified in the DeRe-VM-E Step, namely Entity A and Entity B as exemplified in Fig. \ref{DeRe-VM}, in support of the sparse VBI estimates and forward-backward message passing. Furthermore, this pair of entities need to interact with each other to eliminate the self-reinforcement during the update procedure.
\begin{itemize}
\item \textbf{Entity~A:} By capitalizing sparse VBI \cite{Bayesian-1}, this entity intends to calculate the approximate conditional marginal posteriors of latent variables, i.e., $\psi \left( {{\bm{v}^\theta }} \right)$. We denote $ {\nu _{\chi  \to \alpha _{{n_z}}^\theta }}\left( {\alpha _{{n_z}}^\theta } \right),{n_z} = 1,...,{\tilde N_z} $  by the input of Entity A, which is actually the message passed from Entity B and incorporates the Markov priors of the structured sparsity associated with the sparse vector ${\mathbf{u}}_{{n_y}}^\theta ,{n_y} = 1,...,{N_y}$. The sparse VBI estimates implemented in Entity A will be elaborated on later.

\item \textbf{Entity~B:} Let ${\nu _{f_{{n_z}}^{\bm{\rho }} \to \alpha _{{n_z}}^\theta }}\left(  \cdot  \right)$ denote the message passed from Entity A, which constitutes the input of Entity B. To facilitate exposition, the input and the output of Entity B are defined as $\nu _{{n_z}}^{in} \triangleq {\nu _{f_{{n_z}}^{\bm{\rho }} \to \alpha _{{n_z}}^\theta }}\left(  \cdot  \right)$ and $\nu _{{n_z}}^{out} \triangleq {\nu _{\chi  \to \alpha _{{n_z}}^\theta }}\left(  \cdot  \right),{n_z} = 1,...,{\tilde N_z}$, respectively. In each algorithm iteration, the message $\nu _{{n_z}}^{out}\left( {\alpha _{{n_z}}^\theta } \right) \triangleq {\nu _{\chi  \to \alpha _{{n_z}}^\theta }}\left( {\alpha _{{n_z}}^\theta } \right)$ can be obtained in accordance with the belief propagation rule \cite{TIT-2001,Bayesian-15}
\begin{equation}
\nu _{{n_z}}^{out}\left( {\alpha _{{n_z}}^\theta } \right) \propto \frac{{\psi \left( {\alpha _{{n_z}}^\theta } \right)}}{{\nu _{{n_z}}^{in}\left( {\alpha _{{n_z}}^\theta } \right)}},\forall {n_z},
\end{equation}
which intuitively equals the approximate posterior with respect to  $\alpha _{{n_z}}^\theta $, since $\nu _{{n_z}}^{in}\left( {\alpha _{{n_z}}^\theta } \right)$ plays the role of Entity A’s prior with respect to  $\alpha _{{n_z}}^\theta $. The inference also holds for Entity B's output, and this pair of entities would interact iteratively until their convergence has been achieved.

\end{itemize}
Having established the method for carrying out message exchange between the twin entities, let us move on to an elaboration of the sparse VBI estimation in Entity~A for closed-form updates of the associated posteriors.

\subsubsection{Entity A: Sparse VBI Estimation}\label{sparseVBI}
Firstly, to determine the approximate conditional marginal posterior  $\psi \left( {{\bm{\alpha }^\theta }} \right)$, the sparse VBI estimation can be outlined as a minimization of the Kullback-Leibler divergence (KLD) with respect to $ p\left( {{\bm{v}^\theta }|{{\mathbf{y}}^\theta };\Delta \bm{\theta }} \right) $ and  $ \psi \left( {{\bm{\alpha }^\theta }} \right) $, thus yielding that
\begin{subequations}
\begin{align}\label{KLD}
{\psi ^*}\left( {{\bm{v}^\theta }} \right) &= \mathop {\arg \min }\limits_{\psi \left( {\bm{v}_j^\theta } \right),\Delta \bm{\theta }} \int {\psi \left( {{\bm{v}^\theta }} \right)\ln \frac{{\psi \left( {{\bm{v}^\theta }} \right)}}{{p\left( {{\bm{v}^\theta }|{{\mathbf{y}}^\theta };\Delta \bm{\theta }} \right)}}} \\
s.t. \quad &\psi \left( {{\bm{v}^\theta }} \right) = \prod\limits_{j \in \mathcal{J}} {\psi \left( {\bm{v}_j^\theta } \right)} .
\end{align}
\end{subequations}
The stationary point ${\psi ^*}\left( {\bm{v}_j^\theta } \right)$ to problem (\ref{KLD}) needs to satisfy 
\begin{align}
&{\psi ^*}\left( {\bm{v}_j^\theta } \right) \nonumber\\
& = \mathop {\arg \min }\limits_{\psi \left( {\bm{v}_j^\theta } \right),\Delta \bm{\theta }} \int {\prod\limits_{j \ne l} {{\psi ^*}\left( {\bm{v}_j^\theta } \right)\psi \left( {\bm{v}_l^\theta } \right)\ln \frac{{\prod\nolimits_{j \ne l} {{\psi ^*}\left( {\bm{v}_j^\theta } \right)\psi \left( {\bm{v}_j^\theta } \right)} }}{{p\left( {{\bm{v}^\theta }|{{\mathbf{y}}^\theta };\Delta \bm{\theta }} \right)}}} } d\bm{v}_j^\theta ,
\end{align}
which can be attained via alternating updates of $ \psi \left( {\bm{v}_j^\theta } \right),\forall j \in \mathcal{J} $. Specifically, for given  $ \psi \left( {\bm{v}_l^\theta } \right),l \ne j $, the optimal ${\psi ^*}\left( {\bm{v}_j^\theta } \right)$ that satisfies the KLD minimization in problem (\ref{KLD}) is specified as
\begin{equation}
{\psi ^*}\left( {{\bm{v}}_j^\theta } \right) \propto \exp \left( {{{\left\langle {\ln p\left( {{{\mathbf{y}}^\theta },{{\bm{v}}^\theta };\Delta \bm{\theta }} \right)} \right\rangle }_{\prod\nolimits_{j \ne l} {\psi \left( {{\bm{v}}_l^\theta } \right)} }}} \right),
\end{equation}
where ${\left\langle {f\left( z \right)} \right\rangle _{h\left( z \right)}} = \int {f\left( z \right)} h\left( z \right)dz$.

\begin{algorithm}[t]
	\small
	\caption{Parametric \textbf{De}composition and Compressed \textbf{Re}construction-based \textbf{V}ariational
		Bayesian Inference \textbf{M}essage Passing (DeRe-VM) Algorithm}
	\label{alg}
	\begin{algorithmic}[1]
		\STATE {\bf{\% Parametric Decomposition}}
		\STATE {Decompose 3D AED parameters $\bm{\theta}$, $\bm{\varphi}$, and $\bf{r}$ by constructing their respective covariance matrix according to (\ref{w_theta}), (\ref{w_varphi}), and (\ref{w_r_2})}.
		\STATE {\bf{\% Compressed Reconstruction}}
		\STATE {Initialize the offset vector $\Delta \bm{\theta} = \bf{0} $, transition probability $p_{0\to 1}$, $p_{0\to 1}$, hyper-parameters ${\mathfrak{a}_{{n_y},{n_z}}^\theta }$, ${\mathfrak{b}_{{n_y},{n_z}}^\theta }$, $ {\bar {\mathfrak{a}}_{{n_y},{n_z}}^\theta } $, $ {\bar {\mathfrak{b}}_{{n_y},{n_z}}^\theta } $, $\kappa ^ \theta$, Armijo step size $\iota$, the maximum iteration number $r_ {\max}$, and the convergence threshold $\epsilon$, $ \epsilon^\prime $.}
		\REPEAT
		\STATE {\bf{DeRe-VM-E Step:}}
		\STATE {\bf{\% Entity A: Sparse VBI Estimation}}
		\REPEAT
		\STATE  Update posteriors $\psi \left( {{{\mathbf{u}}^\theta }} \right)$, $\psi \left( {{\bm{\rho }^\theta }} \right)$, and $\psi \left( {{\bm{\alpha }^\theta }} \right)$ in compliance with (\ref{update-u})-(\ref{update-alpha}), respectively.
		\UNTIL The convergence criteria of DiLuS-STPL-E Step is met. 
		\STATE Obtain the message from Entity~A, i.e., $ \nu _{{n_z}}^{in} \left( {{\alpha _{{n_z}}^\theta }} \right)  \triangleq {\nu _{f_{{n_z}}^{\bm{\rho }} \to \alpha _{{n_z}}^\theta }}\left( {\alpha _{{n_z}}^\theta }  \right) $, and then pass it to Entity~B.
		\STATE {\bf{\% Entity B: Forward-backward Message Passing}}
		\STATE  {Update the message $ \nu _{{n_z}}^{out}\left( {\alpha _{{n_z}}^\theta } \right) \triangleq {\nu _{\chi  \to \alpha _{{n_z}}^\theta }}\left( {\alpha _{{n_z}}^\theta } \right)$, and then send it to Entity~A for the next iteration.}
		\STATE \textbf{DeRe-VM-M Step:}
		\STATE Construct the surrogate function $ {\mathcal{F}\left( {\Delta \bm{\theta };\Delta {\bm{\theta }^{\left( r \right)}}} \right)} $ in (\ref{surrogate-function}) using the output of Entity~A in DeRe-VM-E Step, i.e., $ \psi \left( {{\bm{v}}^t} \right) $.
		\STATE Update $ \Delta {\bm{\theta }^{\left( {r + 1} \right)}} $ according to (\ref{update-delta-theta}).
		\UNTIL The convergence criteria $\left\| {\bm{\mu }_{{n_y}}^{\theta ,\left( {r - 1} \right)} - \bm{\mu }_{{n_y}}^{\theta ,\left( r \right)}} \right\| \le \epsilon  $ and $ \left\| {{\mathbf{\Sigma }}_{{n_y}}^{\theta ,\left( {r - 1} \right)} - {\mathbf{\Sigma }}_{{n_y}}^{\theta ,\left( r \right)}}  \right\| \le \epsilon^\prime $ are met or the maximum iteration number $r_{\max}$ is exceeded.
		\STATE Obtain the optimal posterior $ {\psi ^*}\left( {{\bm{v}^\theta }} \right)$ and the offset vector $\Delta \bm{\theta} $. The estimate of $ {\hat \theta _l} $ can be determined by $ {\hat \theta _l} = {\tilde \theta _{{n_{z,q}}}} + \Delta {\theta _{{n_z}}}, q=1,...,L $.
		\STATE {Regarding the estimates of ${\bm{\varphi}}$ and ${\bf{r}}$, repeat \textbf{Step~3-Step~18}, and we just use ${\varphi}$ and ${r}$ in place of the superscript of associated variables, such as $ {\bf{y}}^\varphi $, $ {\bm{v}}^\varphi $, ${\bf{y}}^r $, and $ {\bm{v}}^r $, etc. Next, we obtain two other estimates of $ {\hat \varphi _l}  $ and $\hat r_l$, as given by $ {\hat \varphi _l} = \arcsin \frac{{\sin \left( {{{\tilde \zeta }_{{n_{y,q}}}} + \Delta {\zeta _{{n_y}}}} \right)}}{{\sin {{\hat \theta }_l}}} $, and $ {\hat r_l} = {\tilde r_{{n_{z,q}}}} + \Delta {r_{{n_z}}},q = 1,...,L $.}
		\STATE {Conduct angular index correction in accordance with (\ref{u_n_y_q})-(\ref{correction}).}
		\STATE {Output the index $ \hat{q} $ indicating the $l$th significant path associated 3D AED parameters ($\theta_l$, $\varphi_l$, $r_l$).}
	\end{algorithmic}
\end{algorithm}

Secondly, the posteriors $\psi \left( {{{\mathbf{u}}^\theta }} \right)$,  $ \psi \left( {{\bm{\rho }^\theta }} \right) $ and  $ \psi \left( {{\bm{\alpha }^\theta }} \right) $ can be updated as follows.
\begin{itemize}
\item Update sparse vector $ {{\mathbf{u}}^\theta } $: $ \psi \left( {{{\mathbf{u}}^\theta }} \right) $ can be updated by a complex Gaussian distribution
\begin{equation}\label{update-u}
\psi \left( {{{\mathbf{u}}^\theta }} \right) = \prod\limits_{{n_y} = 1}^{{N_y}} {\mathcal{C}\mathcal{N}\left( {{\mathbf{u}}_{{n_y}}^\theta ;\bm{\mu }_{{n_y}}^\theta ,{\mathbf{\Sigma }}_{{n_y}}^\theta } \right)} .
\end{equation}
$ {{\mathbf{\Sigma }}_{{n_y}}^\theta } $ and $ \bm{\mu }_{{n_y}}^\theta $ can be calculated by
\begin{align}
	&\resizebox{0.9\hsize}{!}{$
		{{\mathbf{\Sigma }}_{{n_y}}^\theta } 
		= {\left( {{\text{diag}}\left( {{\rho _{{n_y},1}},...,{\rho _{{n_y},{{\tilde N}_z}}}} \right) + {\kappa ^\theta }{\mathbf{\Psi }}_{{n_y}}^\theta {{\left( {\Delta \bm{\theta }} \right)}^H}{\mathbf{\Psi }}_{{n_y}}^\theta \left( {\Delta \bm{\theta }} \right)} \right)^{ - 1}},
		$}\nonumber\\
	&\quad \ \  \resizebox{0.45\hsize}{!}{$
		={\mathbf{\Lambda }}_{{n_y}}^\theta  - \left\langle {{\kappa ^\theta }} \right\rangle {\mathbf{\Lambda }}_{{n_y}}^\theta {\mathbf{\Psi }}_{{n_y}}^\theta {\left( {\Delta \bm{\theta }} \right)^H} 
		$} \nonumber\\ 
	&\quad \ \ \resizebox{0.85\hsize}{!}{$
		\times{\left( {{\mathbf{I}} + \left\langle {{\kappa ^\theta }} \right\rangle {\mathbf{\Psi }}_{{n_y}}^\theta \left( {\Delta \bm{\theta }} \right){\mathbf{\Lambda }}_{{n_y}}^\theta {\mathbf{\Psi }}_{{n_y}}^\theta {{\left( {\Delta \bm{\theta }} \right)}^H}} \right)^{ - 1}}{\mathbf{\Psi }}_{{n_y}}^\theta \left( {\Delta \bm{\theta }} \right){\mathbf{\Lambda }}_{{n_y}}^\theta ,$}
	\label{inversion} 	 
\end{align}
\begin{align}
	 \bm{\mu }_{{n_y}}^\theta  = \left\langle \kappa^\theta  \right\rangle {\mathbf{\Sigma }}_{{n_y}}^\theta {\mathbf{\Psi }}_{{n_y}}^\theta {\left( {\Delta \bm{\theta }} \right)^H}{\mathbf{y}}_{{n_y}}^\theta ,
\end{align}
where $ \left\langle {{\rho _{{n_y},{n_z}}}} \right\rangle  = \frac{{\tilde {\mathfrak{a}}_{{n_y},{n_z}}^\theta }}{{\tilde {\mathfrak{b}}_{{n_y},{n_z}}^\theta }} $, $ \left\langle \kappa^\theta  \right\rangle  = \frac{{\mathfrak{a}_\kappa ^\theta }}{{\mathfrak{b}_\kappa ^\theta }}$, and $ {\mathbf{\Lambda }}_{{n_y}}^\theta  = {\text{diag}}\left\{ {\frac{{\tilde {\mathfrak{b}}_{{n_y},1}^\theta }}{{\tilde {\mathfrak{a}}_{{n_y},1}^\theta }},...,\frac{{\tilde {\mathfrak{b}}_{{n_y},{{\tilde N}_z}}^\theta }}{{\tilde {\mathfrak{a}}_{{n_y},{{\tilde N}_z}}^\theta }}} \right\} $.

\item Update precision vector $ {\bm{\rho }^\theta } $: $ \psi \left( {{\bm{\rho }^\theta }} \right) $ is given by  \begin{equation}\label{update-rho}
\psi \left( {{\bm{\rho }^\theta }} \right) = \prod\limits_{{n_y} = 1}^{{N_y}} {\prod\limits_{{n_z} = 1}^{{{\tilde N}_z}} {\Gamma \left( {\rho _{{n_y},{n_z}}^\theta ;\tilde {\mathfrak{a}}_{{n_y},{n_z}}^\theta ,\tilde {\mathfrak{b}}_{{n_y},{n_z}}^\theta } \right)} } ,
\end{equation}
where $\tilde {\mathfrak{a}}_{{n_y},{n_z}}^\theta  = \tilde \nu _{{n_z}}^{out}\mathfrak{a}_{{n_y},{n_z}}^\theta  + \left( {1 - \tilde \nu _{{n_z}}^{out}} \right)\bar {\mathfrak{a}}_{{n_y},{n_z}}^\theta  + 1$, $ \tilde {\mathfrak{b}}_{{n_y},{n_z}}^\theta  = \tilde \nu _{{n_z}}^{out}\mathfrak{b}_{{n_y},{n_z}}^\theta  + \left( {1 - \tilde \nu _{{n_z}}^{out}} \right)\bar {\mathfrak{b}}_{{n_y},{n_z}}^\theta  + {\left| {\bm{\mu }_{{n_y}}^\theta } \right|^2} + \Sigma _{{n_y},{n_z}}^\theta  $, and $\tilde \nu _{{n_z}}^{out}$ represents the posterior probability of $ \alpha _{n_z} ^\theta =1$, also known as the message passed from Entity~A.

\item Update support vector $ {\bm{\alpha }^\theta } $: We obtain the posterior $\psi \left( {{\bm{\alpha }^\theta }} \right)$ in the form of
\begin{equation}\label{update-alpha}
\psi \left( {{\bm{\alpha }^\theta }} \right) = \prod\limits_{{n_z} = 1}^{{{\tilde N}_z}} {\left[ {\tilde \nu _{{n_z}}^{out}\delta \left( {\alpha _{{n_z}}^\theta  - 1} \right) + \left( {1 - \tilde \nu _{{n_z}}^{out}} \right)\delta \left( {\alpha _{{n_z}}^\theta } \right)} \right]} .
\end{equation}

\end{itemize}

Likewise, the foregoing procedures are also applicable to both $\bm{\varphi}$ and $\bf{r}$ for obtaining their CS-based solutions. When establishing the layered processing probability model and conducting sparse estimation techniques, we just use ${\varphi}$ and ${r}$ in place of the superscript of associated variables, e.g., $ {\bf{y}}^\varphi $ and ${\bf{y}}^r $, etc., and the corresponding phases have been omitted for conciseness. In brief, the proposed DeRe-VM algorithm have been summarized in Algorithm~\ref{alg}.

\subsection{Computational Complexity}
Following the overall flow of DDS-VBIMP summarized in Algorithm 1, the main computational burden lies in the updates of  $ \psi \left( {{{\mathbf{u}}^\theta }} \right) $,  $ \psi \left( {{{\bm{\rho }}^\theta }} \right) $, and  $ \psi \left( {{{\bm{\alpha }}^\theta }} \right) $ in Step 9 using (\ref{update-u})-(\ref{update-alpha}), respectively. The total number of multiplications to update $ \psi \left( {{{\mathbf{u}}^\theta }} \right) $ is  $ {N_y}\left( {3{N_{{\text{RF}}}}\tilde N_z^2 + 2{{\tilde N}_z}N_{{\text{RF}}}^2} \right) $. Upon the utilization of the matrix inversion lemma on (\ref{inversion}) in order to lower complexity, the computational complexity with regard to the matrix inversion in (\ref{inversion}) is $ \mathcal{O}\left( {{N_y}N_{{\text{RF}}}^3} \right) $. Thus, the total computational complexity order of the proposed DeRe-VM algorithm is $\mathcal{O}\left( {{N_{{\text{RF}}}}{N_y}\tilde N_z^2 + {N_y}{{\tilde N}_z}N_{{\text{RF}}}^2} \right)$ per iteration, concerning that it typically holds that $ {N_y},{{\tilde N}_z} \gg {N_{{\text{RF}}}}  $. This implies that for an $ N $-element UPA ($N= N_y\times N_z$), the complexity of the proposed DeRe-VM algorithm just scales proportionally with the number of antenna elements raised to the power of 1.5.
Additionally, our proposed DeRe-VM algorithm functions within the off-grid basis and is capable of eliminating on-grid errors, sharing the similarities with the methodology presented in \cite{XLM-1}, namely polar-domain simultaneous gridless weighted (P-SIGW). Albeit the evident superiority of P-SIGW, there are two issues that should be taken with a grain of salt: i) P-SIGW fails to capture the coherence between the 2D elevation-azimuth angle pair $(\bm{\theta}, \bm{\varphi})$ and the distance $\mathbf{r}$ for a robust recovery of 3D AED parameters, and ii) the computational complexity of P-SIGW increases quadratically with the number of antenna elements, making it less favorable for a large $N$. In the section that follows, a comprehensive performance evaluation will be illustrated, showcasing their distinct advantages in relation to various cases.

\section{Simulation Results}
\subsection{Configuration}
In this section, we conduct performance evaluation of the proposed DeRe-VM algorithm for channel estimation in the holographic MIMO system of interest. 
The antenna elements at the BS are equipped in the form of ${N_y} = 129$ and $N_z = 65$. The considered holographic MIMO system operates at 30 GHz with a bandwidth of 100~MHz. It is assumed that the number of paths is $L=3$. The simulation results are averaged over 500~random channel realizations. For each realization, the distance between the BS and the user/scatterer is sampled from a uniform distribution ${r_l} \sim \mathcal{U}\left( {{\mathfrak{D}_{\min }},0.5{\mathfrak{D}_{\max }}} \right),l = 1,...,L$, and based on the fixed distance, the associated angular parameters of the LoS path are randomly sampled from the uniform distributions ${\theta _l} \sim \mathcal{U}\left( {0,\pi } \right)$  and  ${\varphi _l} \sim \mathcal{U}\left( { - \pi /2,\pi /2} \right),l = 1,...,L$, respectively.  Furthermore, the complex gain of each NLoS path is modeled to follow a complex Gaussian distribution with zero mean and variance ${\sigma ^{{\text{NL}}}}$, i.e.,  $ \mathcal{C}\mathcal{N}\left( {0,{\sigma ^{{\text{NL}}}}} \right) $, in which ${\sigma ^{{\text{NL}}}}$ is 20~dBm weaker than that of the LoS paths \cite{RIS-LOC-10}. In the process of the CS reconstruction, the number of the uniform grid is set to ${\tilde N_z} = {N_z}$ for $\bm{\theta}$, and ${\tilde N_y} = {N_y}$ for $\bm{\varphi}$, respectively. 
We concentrate on the normalized mean square error (NMSE) performance of the auto-correlation matrix $\bf{R}$, which can be reconstructed by ${\mathbf{R}} \triangleq \mathbb{E}\left\{ {{\mathbf{h}}{{\mathbf{h}}^H}} \right\}$ after we obtain the estimates of the 3D AED parameters $\hat {\bm{ \theta }}$,  $\hat {\bm{ \varphi }}$, and  ${\hat {\mathbf{r}}}$. The NMSE is defined as  $ \mathbb{E}\left[ {\left\| {{\mathbf{R}} - {\mathbf{\hat R}}} \right\|_F^2/\left\| {\mathbf{R}} \right\|_F^2} \right] $. Other key system parameters are set as follows unless otherwise specified: $ {N_{{\text{RF}}}} = 16 $, $ P=32 $, $P_p = 33$~dBm, $ {\beta _\Delta } = 1.2 $, $ \frac{{{p_{0 \to 1}}}}{{{p_{0 \to 1}} + {p_{1 \to 0}}}} = 0.06 $, $ \mathfrak{a}_{{n_y},{n_z}}^i = \bar {\mathfrak{a}}_{{n_y},{n_z}}^i = \mathfrak{b}_{{n_y},{n_z}}^i = 0.99 $, $\bar {\mathfrak{b}}_{{n_y},{n_z}}^i = 0.01$, $i \in \left\{ {\theta ,\varphi ,r} \right\}$.

\subsection{Convergence Behavior}

\begin{figure}
	\centering
	\includegraphics[width=0.49\textwidth]{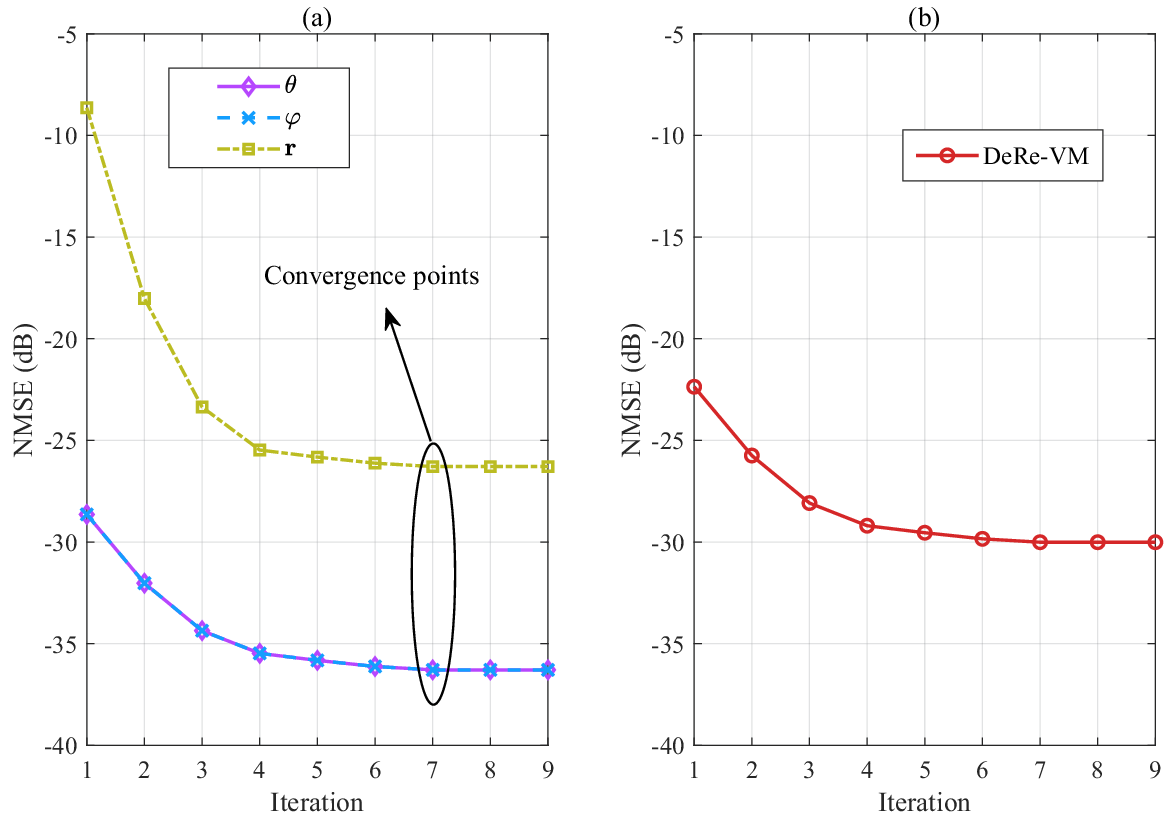}
	\caption{Convergence behavior of DeRe-VM. (a) NMSE performance of 3D AED parameters $\bm{ \theta }$, $\bm{ \varphi }$, and $\bf{r}$. (b) NMSE performance of the auto-correlation matrix $\mathbf{R}$.} \label{Converge}
\end{figure}

We demonstrate the convergence behavior of the proposed DeRe-VM algorithm by plotting the NMSE curves of 3D AED parameters $\bm{ \theta }$, $\bm{ \varphi }$, and $\bf{r}$ in Fig.~\ref{Converge}(a), which is calculated by $ \mathbb{E}\left[ {\left\| {{\mathbf{x}} - {\mathbf{\hat x}}} \right\|_2^2/\left\| {\mathbf{x}} \right\|_2^2} \right], \forall  \mathbf{x} \in \left\lbrace  {\bm{ \theta }, \bm{ \varphi },  {\mathbf{r}}} \right\rbrace $~\cite{Li2017}. By contrast, Fig.~\ref{Converge}(b) shows the NMSE performance of the auto-correlation matrix, obtained by $ \mathbb{E}\left[ {\left\| {{\mathbf{R}} - {\mathbf{\hat R}}} \right\|_F^2/\left\| {\mathbf{R}} \right\|_F^2} \right] $. A steep decline can be spotted in the first few iterations, followed by a gradual settling at a value after around seven iterations for both scenarios. 
The NMSE gap in accuracy between $\mathbf{r}$ and $\bm{\theta} (\bm{ \varphi })$ can be attributed to the presence of a residual correlation inside the distance dictionary matrix. Albeit this non-orthogonality, DeRe-VM still exhibits favorable estimation precision in terms of $\mathbf{R}$'s NMSE performance.

\begin{figure}
	\centering
	\includegraphics[width=0.48\textwidth]{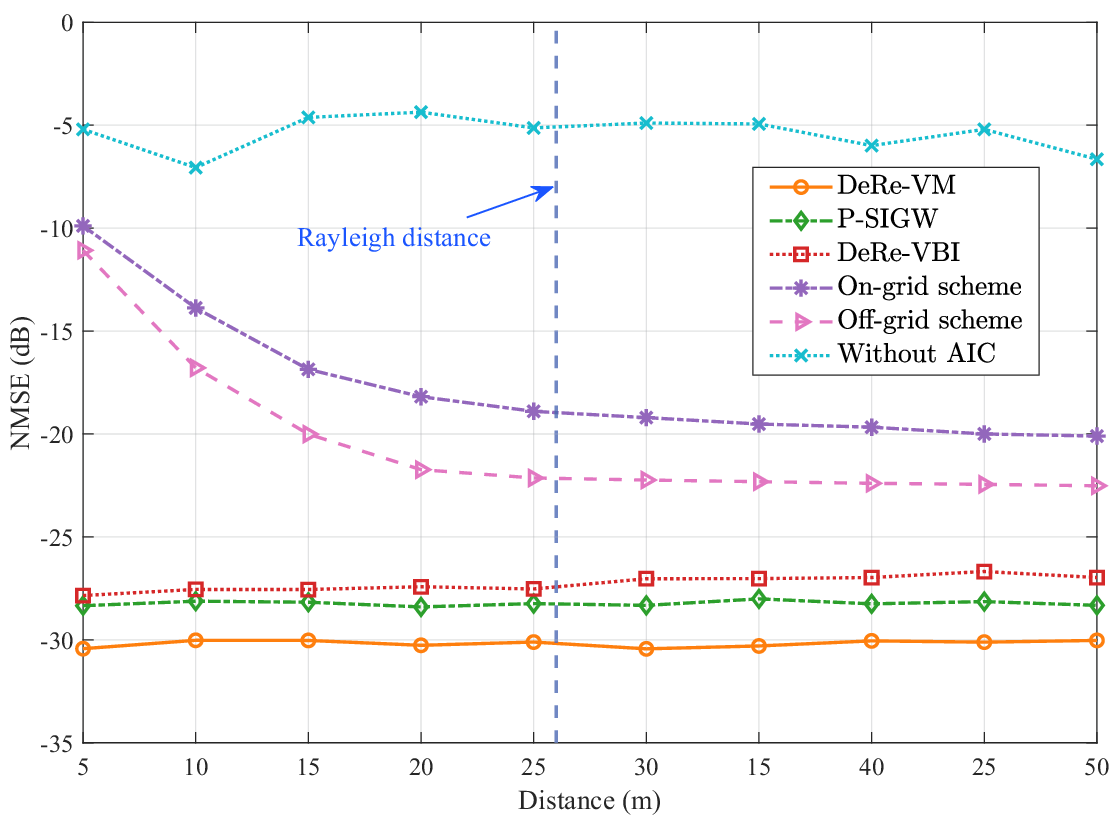}
	\caption{NMSE performance versus the distance. } \label{NMSEvsDist}
\end{figure}


\subsection{NMSE Performance}
For comparison, the exploited benchmark schemes are listed as follows:
\begin{itemize}
\item \textbf{P-SIGW} \cite{XLM-1}: The off-grid polar-domain simultaneous iterative gridless weighted (P-SIGW) algorithm exploits different transform matrix constructed for the independent decomposition of angle and distance parameters.	
	
\item \textbf{DeRe-VBI} \cite{Bayesian-1}: In lieu of Markov prior in (\ref{Markov}), DeRe-VBI assumes the independent and identically distributed (i.i.d.) sparse support vector priors. 

\item \textbf{Off-grid scheme} \cite{chen-twc3}: The holographic MIMO channel is treated in the angular domain using VBI approach. 

\item \textbf{On-grid scheme} \cite{RIS-CE-12}: The holographic MIMO channel is treated in the angular domain using a discrete Fourier transform (DFT)-based on-grid method proposed in~\cite{RIS-CE-12}. 

\item \textbf{Without AIC}: The angular index correction in Step~20 of Algorithm~\ref{alg} is removed.


\end{itemize}

\begin{figure}
	\subfigure[]{
		\begin{minipage}[t]{0.49\textwidth}
			\centering
			\includegraphics[width=\textwidth]{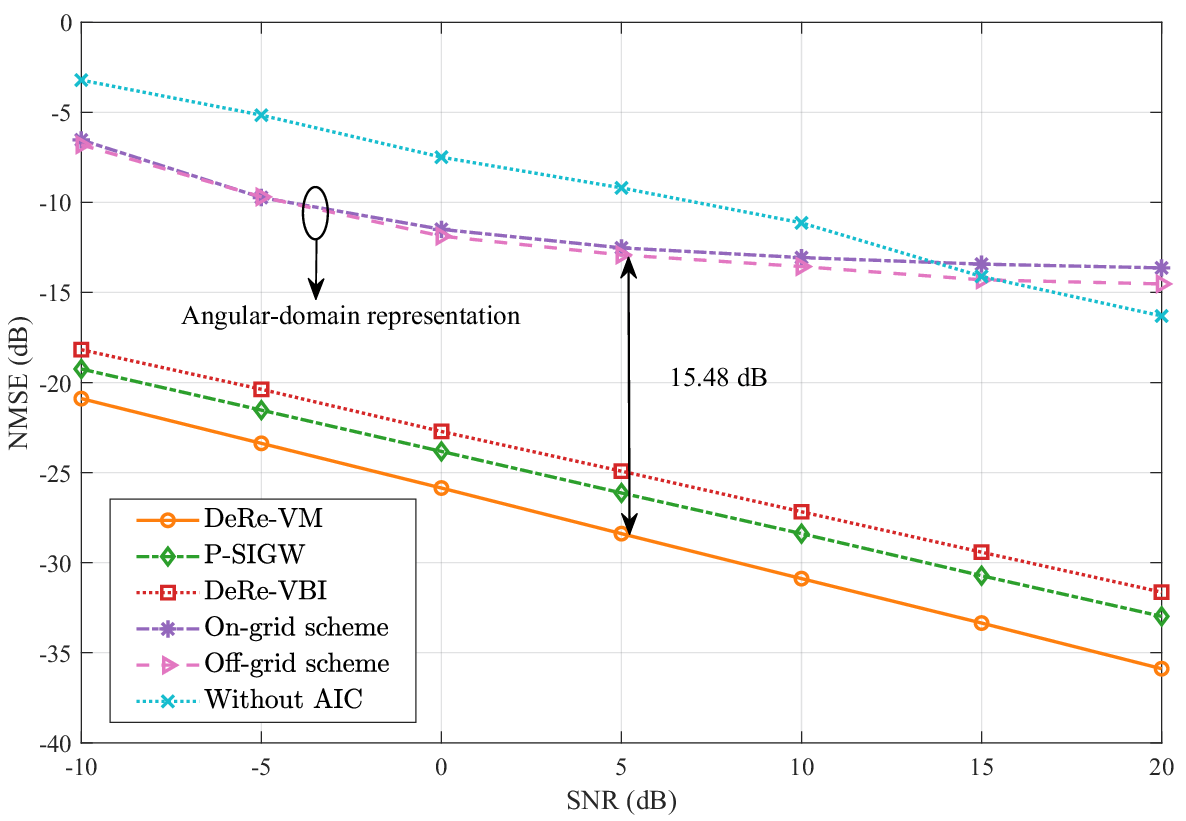}
			\label{NMSEvsSNR_a}
	\end{minipage}}
	\subfigure[]{
		\begin{minipage}[t]{0.49\textwidth}
			\centering
			\includegraphics[width=\textwidth]{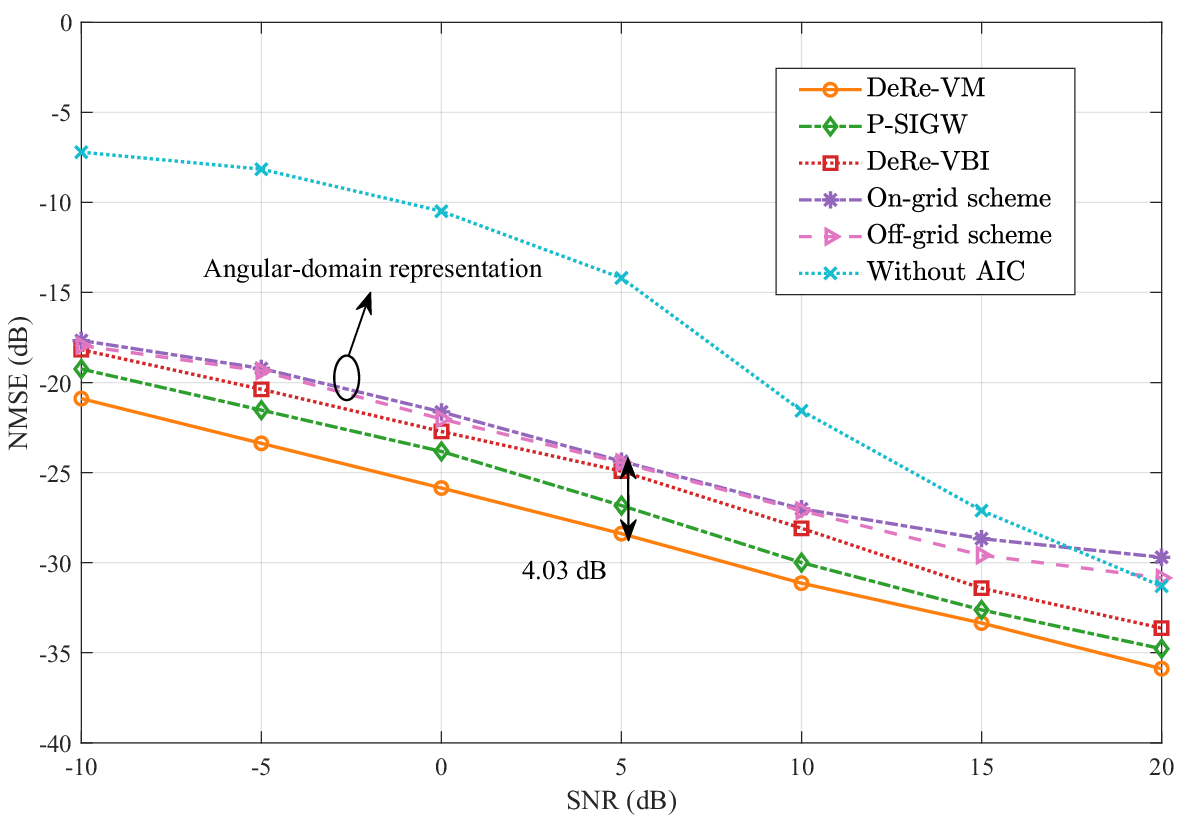}
			\label{NMSEvsSNR_b}
	\end{minipage}}
	\caption{NMSE performance versus SNR with different BS-user distances. (a) The BS-user distance is taken from $r_l \sim \mathcal{U}\left( {1\,{\text{m}},20\,{\text{m}}} \right)$ for the near-field region. (b) The BS-user distance is taken from $r_l \sim \mathcal{U}\left( {30 \, {\text{m}},50\, {\text{m}}} \right)$ for the far-field region.}\label{NMSEvsSNR}
\end{figure}

\begin{figure}
	\subfigure[]{
		\begin{minipage}[t]{0.49\textwidth}
			\centering
			\includegraphics[width=\textwidth]{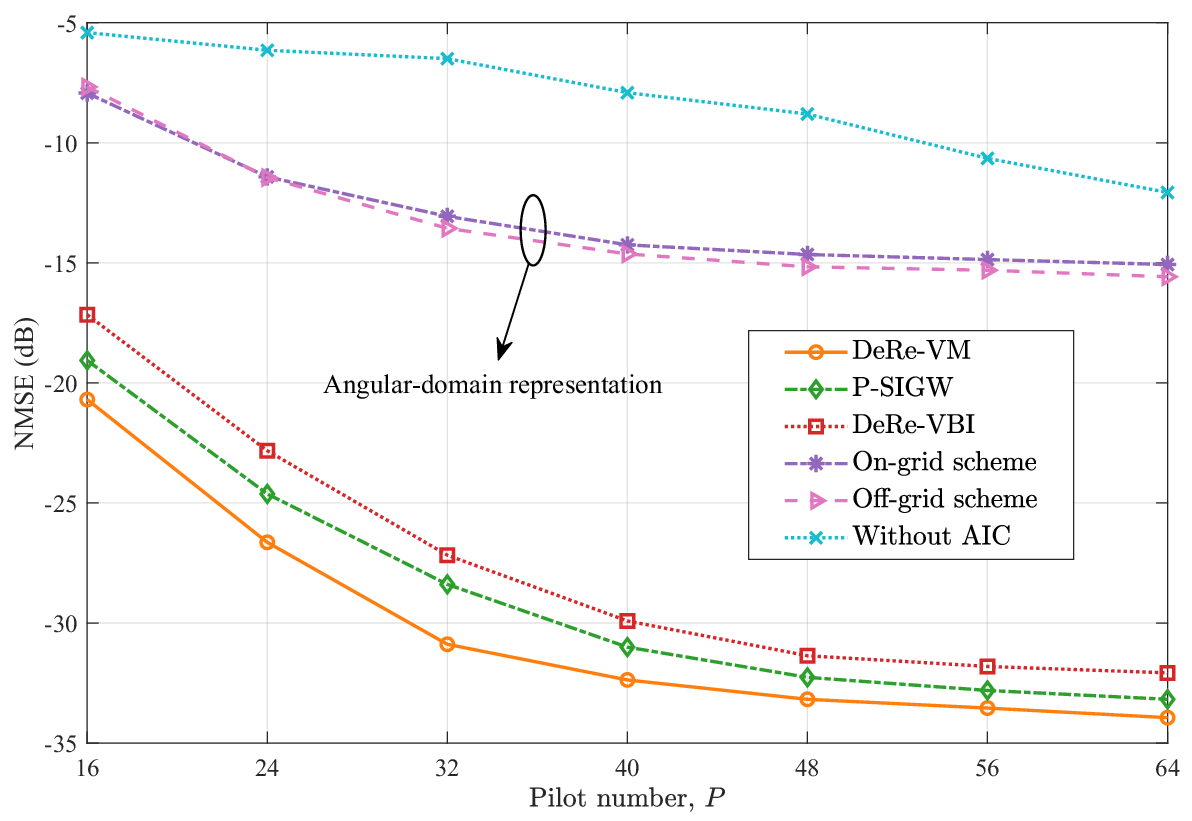}
			\label{NMSEvsPilot_a}
	\end{minipage}}
	\subfigure[]{
		\begin{minipage}[t]{0.49\textwidth}
			\centering
			\includegraphics[width=\textwidth]{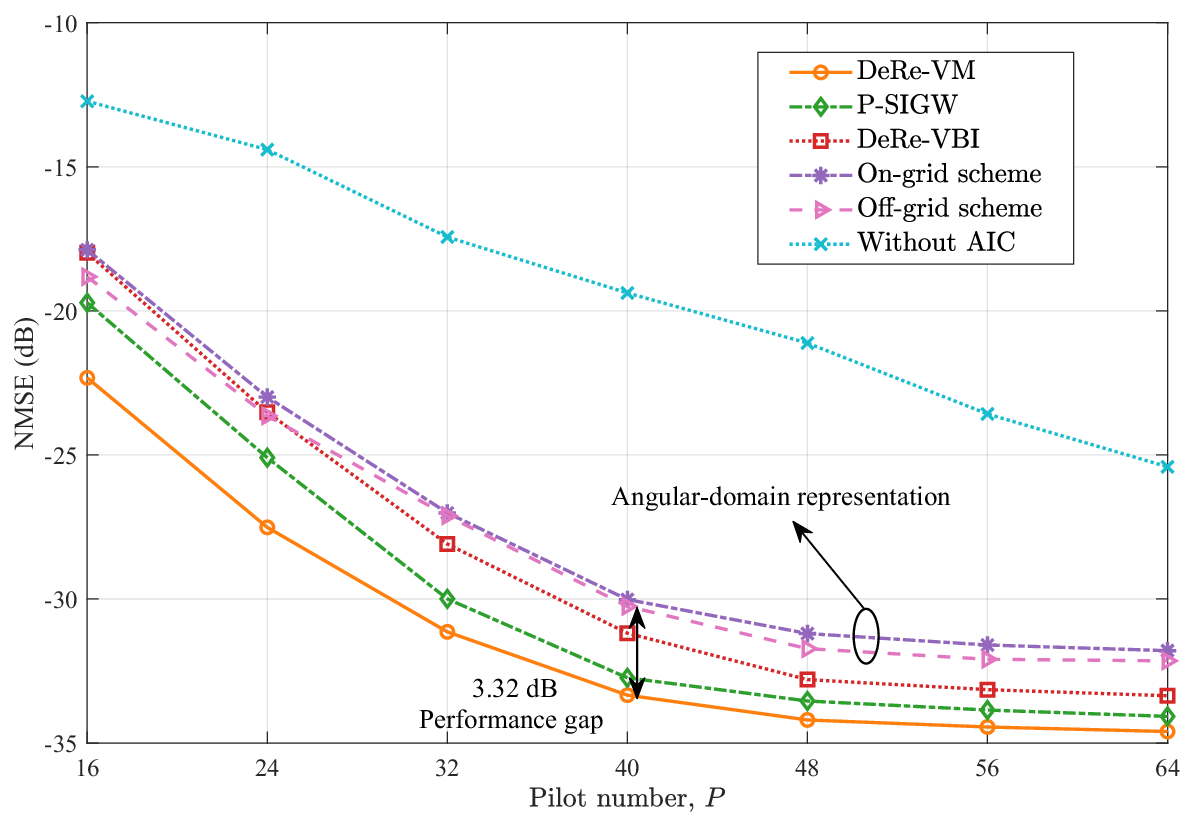}
			\label{NMSEvsPilot_b}
	\end{minipage}}
	\caption{NMSE performance versus the pilot number $P$ with different BS-user distances. (a) The BS-user distance is taken from $r_l \sim \mathcal{U}\left( {1\,{\text{m}},20\,{\text{m}}} \right)$ for the near-field region. (b) The BS-user distance is taken from $r_l \sim \mathcal{U}\left( {30 \, {\text{m}},50\, {\text{m}}} \right)$ for the far-field region.}\label{NMSEvsPilot}
\end{figure}

We begin the NMSE performance evaluation by investigating their corresponding curves versus the BS-to-user distance ranging from 5 m to 50 m. In accordance with the carrier frequency of 30 GHz and the number of antenna elements of ${N_y} = 129$ and $N_z = 65$, the Rayleigh distance can be determined to be 26~meters. As shown in Fig.~\ref{NMSEvsDist}, the proposed DeRe-VM algorithm consistently reveals the best NMSE performance, whereas that of the scheme Without AIC swings erratically as the distance increases from the near-field to the far-field.  The poor NMSE performance attained by the scheme Without AIC results from the ambiguity of $\bf{R}$ imposed by the thoughtlessly stitching of the 3D AED parameters, as the angular index misinterpretation frequently occurs. Furthermore, a NMSE performance leveling off is spotted with respect to the schemes incorporating distance estimation, e.g., DeRe-VM, P-SIGW, and DeRe-VBI, regardless of fluctuations in distance. In contrast, both on-grid and off-grid schemes exhibit poor estimating performance, particularly in the near-field region. This is primarily due to the fact that distance-estimation schemes are capable of structuring distance grids in a sense that accommodates to various channel environments. More precisely, when doing gridding over the distance $\bf{r}$, a uniform grid of discrete points embraces the far-field case, i.e., $n_z = 0$ in (\ref{r_grid}), allowing the far-field information to be retrieved from the $\bf{r}$-based covariance matrix. Therefore, the proposed DeRe-VM reveals significant robustness in accommodating diverse channel conditions at varying distances in the context of a holographic MIMO system.

In Fig.~\ref{NMSEvsSNR}, we investigate the impact of varying SNR on the NMSE performance for different BS-to-user distance configurations, where the 26-meter Rayleigh distance is taken as the dividing line for the near-field and far-field regions.
Firstly, as expected, an increase in SNR boosts the NMSE performance and the proposed DeRe-VM algorithm is superior to the benchmark schemes adopted, regardless of the distance configurations. This is made possible by the fact that DeRe-VM offers the capability of capturing, via their independent decompositions, potential structured sparsities in relation to 3D AED parameters in holographic MIMO channels. On the other hand, the scatterers in the physical environment can be characterized by Markov prior for the channel support $\bm{\alpha} ^ \theta$, which also explains the reason why DeRe-VM is superior to DeRe-VBI. Secondly, the NMSE performance achieved by on-grid and off-grid schemes significantly enhances when the examined settings shift to the far-field case. For instance, as SNR = 5~dB, the performance gap resulted from DeRe-VM and off-grid schemes falls from 15.48~dB in the near-field case to 4.03~dB in the far-field case. Additionally, one interesting finding is that in the far-field configuration, higher SNR draws the NMSE performance achieved by Without AIC to a great level, in stark contrast to the pale efforts made by the increased SNR for the marginal improvement of NMSE in the near-field context. This is because a high SNR can compensate for the erroneous estimation caused by the absence of the angular index correction when just estimating the 2D azimuth-elevation angle pair.

\begin{figure}
	\centering
	\includegraphics[width=0.5\textwidth]{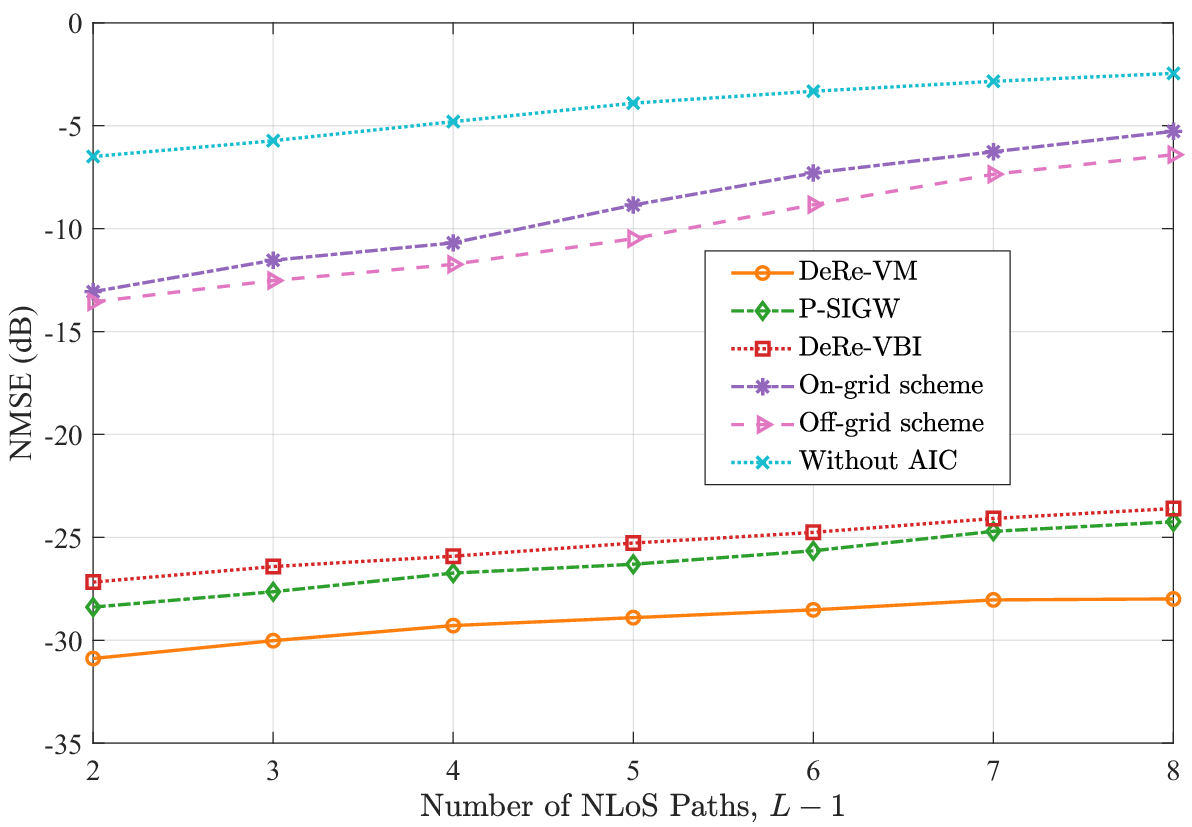}
	\caption{NMSE performance versus the number of NLoS  paths ($L-1$).} \label{NMSEvsNLOS}
\end{figure}

Fig.~\ref{NMSEvsPilot} plots the NMSE curves in relation to the pilot length $P$ for various distance cases, as in Fig.~\ref{NMSEvsSNR}. With more pilots inserted, holographic MIMO channels can be better characterized, resulting in a diffing degree of improvement for NMSE with various schemes. To be specific, regarding ${r_l} \sim \mathcal{U}\left( {{\text{1 m}},20{\text{ m}}} \right)$ showcased in Fig.~\ref{NMSEvsPilot_a}, an increment in $P$ boost NMSE performance marginally in the near-field case while considerably in the far-field case. This can be attributed to the fact that on- and off-grid schemes based on the angular-domain representation suffer from a performance bottleneck induced by a severe energy spread and erroneous detection of significant paths. Interestingly, both cases demonstrate a slight decline in NMSE when $ P > 40$, in contrast to a precipitous drop when $P < 40$, which implies a potential compromise between the pilot number and NMSE, that is, favorable NMSE performance can be achieved given an appropriate $P$, while a large $P$ may incur heavy pilot overhead for systems. Additionally, the performance gap can still be spotted in Fig.~\ref{NMSEvsPilot_b}, e.g., 3.32 dB when $P=40$, between DeRe-VM and On-grid schemes, owing to a more accurate Fresnel approximation in (\ref{Fresnel})  than its far-field counterpart.

Fig.~\ref{NMSEvsNLOS} demonstrates the variation of NMSE performance with respect to the number of NLoS paths among different schemes. Since $L=1$ represents the LoS path, the number of NLoS paths is denoted by $L-1$. As spotted in Fig.~\ref{NMSEvsNLOS}, the NMSE performance achieved by all the schemes being considered exhibits a slight degradation with an increased number of NLoS paths, albeit to different degrees. As the number of NLoS paths increases, an augmentation occurs in the quantity of non-zero entries within each sparse vector. This without a doubt increases the dimensionality of the solution space, rendering the CS-based algorithm more susceptible to becoming ensnared in local optima in the pursuit of the optimal value. As a result, figuring out the LoS path becomes more challenging. Fortunately, due to the ``turbo” nature of DeRe-VM, with which the updates of parameters and posteriors mutually reinforce each other, it can be ensured that the algorithm will eventually converge to a stationary point~\cite{LOC-1,chen-jsac2}.

\begin{figure}
	\centering
	\includegraphics[width=0.5\textwidth]{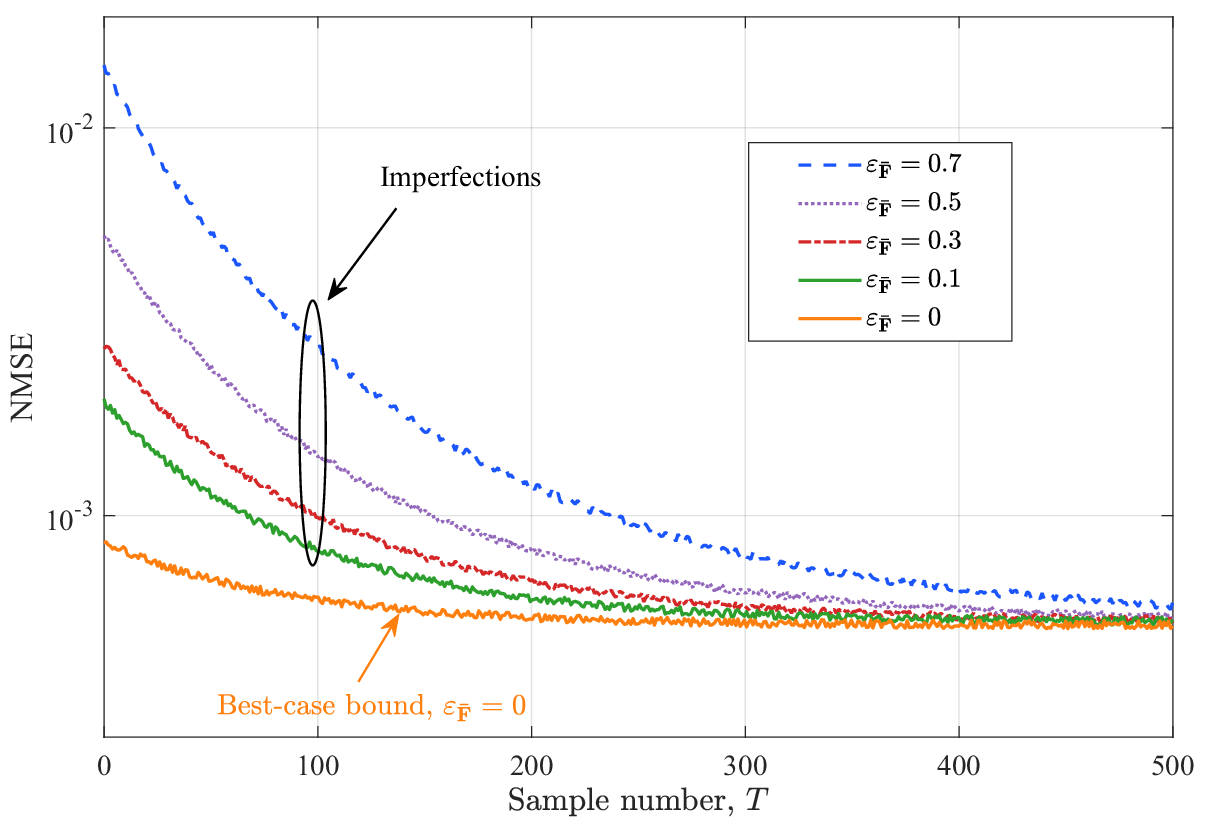}
	\caption{NMSE performance at different error ratios ${\varepsilon _{{\mathbf{\bar F}}}}$.} \label{error_ratio}
\end{figure}

To evaluate the robustness of the proposed DeRe framework, Fig.~\ref{error_ratio} plots the NMSE curves versus the sample number $T$. In conjunction with (\ref{mapping_y_theta2}), we define ${\varepsilon _{{\mathbf{\bar F}}}} = \sqrt {\mathbb{E}\left\{ {{{\left\| {\Delta {\mathbf{\bar F}}} \right\|}^2}} \right\}/\mathbb{E}\left\{ {{{\left\| {{\mathbf{\bar F}} + \Delta {\mathbf{\bar F}}} \right\|}^2}} \right\}} $ as the error ratio to quantify the level of hardware imperfections induced by for example, the limited resolution of phase shifters and amplitude controllers. As it transpires, the NMSE performance is expected to improve with an increase in the number of samples. An eye-catching observation pertains to the existence of gaps in NMSE performances across various ${\varepsilon _{{\mathbf{\bar F}}}}$ in the presence of a small number of training samples, while the NMSE performance achieved by all cases, as $T$ grows larger, tends to approach the best-case bound. This verifies that the DeRe framework has the capability to learn the statistical properties of the 3D AED parameters by harnessing the samples adequately despite the presence of hardware imperfections, showcasing the great robustness of the proposed DeRe framework.

\section{Conclusion}
In this work, we have investigated the channel estimation for holographic MIMO systems in face of a few non-trivial challenges posed by the holographic MIMO nature. A sophisticated DeRe framework is devised for deep decouplings of the 3D AED parameters, followed by an efficient DeRe-VM algorithm for resolving the resultant CS problems. We evince that the implementation of the proposed DeRe framework facilitates sharp detection of the parameters for each dimension, but it may adversely results in misinterpretation of the angular index if these quantities obtained are not properly structured together. As a remedy, angular index correction is carried out for the accurate estimation. Our simulation results substantiate that the proposed DeRe-VM algorithm exhibits great robustness to the distance fluctuations in the holographic MIMO system of interest, irrespective of whether the scenario involves near-field or far-field. In contrast to the conventional far-field solutions based on the angular domain, the proposed DeRe-VM paradigm still shows slightly improved NMSE performance owing to its employment of a more accurate Fresnel approximation, without noticeable increase in sparsifying complexity. 


%

\ifCLASSOPTIONcaptionsoff
  \newpage
\fi


\bibliographystyle{IEEEtran}
\bibliography{ref_Holo}

\begin{IEEEbiography}[{\includegraphics[width=1in,height=1.25in,clip,keepaspectratio]{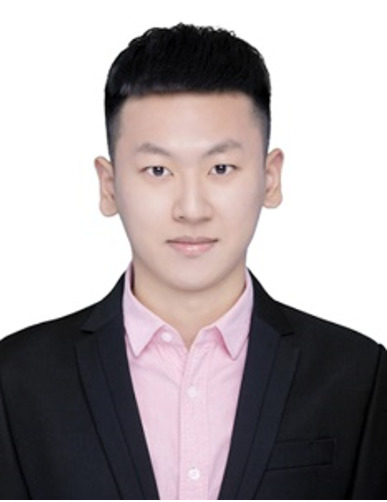}}]{Yuanbin Chen}
	received the B.S. degree in communications engineering from the Beijing Jiaotong University, Beijing, China, in 2019. He is currently pursuing the Ph.D. degree in information and communication systems with the State Key Laboratory of Networking and Switching Technology, Beijing University of Posts and Telecommunications. His current research interests are in the area of holographic MIMO, reconfigurable intelligent surface (RIS), and radio resource management (RRM) in future wireless networks. He was the recipient of the National Scholarship in 2020 and 2022.
\end{IEEEbiography}

\begin{IEEEbiography}[{\includegraphics[width=1in,height=1.25in,clip,keepaspectratio]{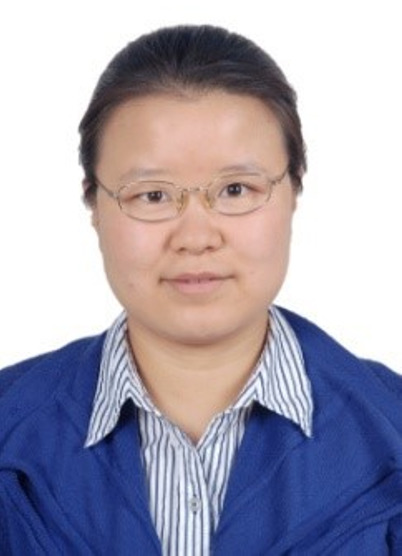}}]{Ying Wang}
	(IEEE Member) received the Ph.D. degree in circuits and systems from the Beijing University of Posts and Telecommunications (BUPT), Beijing, China, in 2003. 
	In 2004, she was invited to work as a Visiting Researcher with the Communications Research Laboratory (renamed NiCT from 2004), Yokosuka, Japan. She was a Research Associate with the University of Hong Kong, Hong Kong, in 2005. She is currently a Professor with BUPT and the Director of the Radio Resource Management Laboratory, Wireless Technology Innovation Institute, BUPT. Her research interests are in the area of the cooperative and cognitive systems, radio resource management, and mobility management in 5G systems. She is active in standardization activities of 3GPP and ITU. She took part in performance evaluation work of the Chinese Evaluation Group, as a Representative of BUPT. She was a recipient of first prizes of the Scientific and Technological Progress Award by the China Institute of Communications in 2006 and 2009, respectively, and a second prize of the National Scientific and Technological Progress Award in 2008. She was also selected in the New Star Program of Beijing Science and Technology Committee and the New Century Excellent Talents in University, Ministry of Education, in 2007 and 2009, respectively. She has authored over 100 papers in international journals and conferences proceedings.
\end{IEEEbiography}

\begin{IEEEbiography}[{\includegraphics[width=1in,height=1.25in,clip,keepaspectratio]{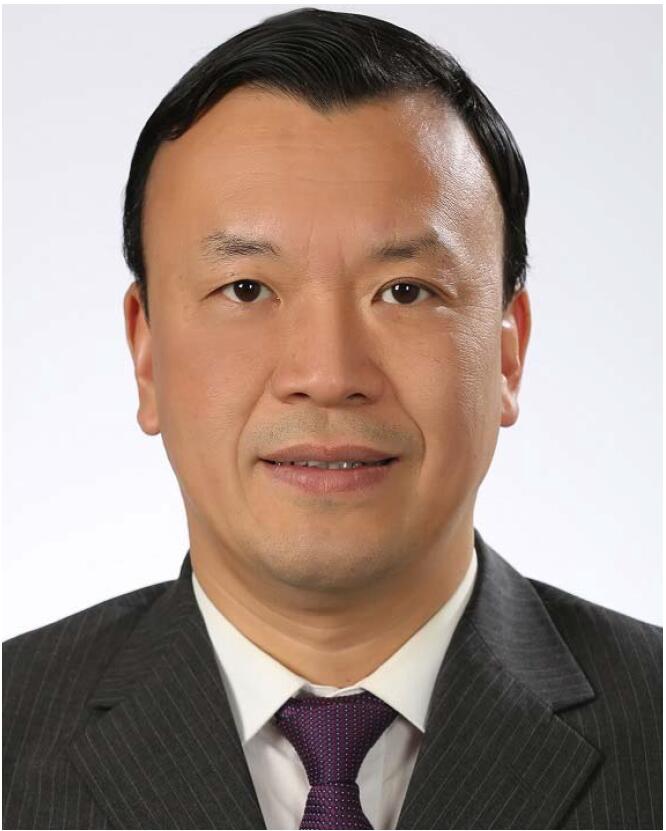}}]{Zhaocheng Wang} (IEEE Fellow) received the B.S., M.S., and Ph.D. degrees from Tsinghua University, in 1991, 1993, and 1996, respectively. 
	
	From 1996 to 1997, he was a Post-Doctoral Fellow with Nanyang Technological University, Singapore. From 1997 to 1999, he was a Research Engineer/ Senior Engineer with the OKI Techno Centre (Singapore) Pte. Ltd., Singapore. From 1999 to 2009, he was a Senior Engineer/Principal Engineer with Sony Deutschland GmbH, Germany. Since 2009, he has been a Professor with the Department of Electronic Engineering, Tsinghua University, where he is currently the Director of the Broadband Communication Key Laboratory, Beijing National Research Center for Information Science and Technology (BNRist). He has authored or coauthored two books, which have been selected by IEEE Series on Digital and Mobile Communication and published by Wiley-IEEE Press. He has authored/coauthored more than 200 peer-reviewed journal articles. He holds 60 U.S./EU granted patents (23 of them as the first inventor). His research interests include wireless communications, millimeter wave communications, and optical wireless communications. He is a fellow of the Institution of Engineering and Technology. He was a recipient of the ICC2013 Best Paper Award, the OECC2015 Best Student Paper Award, the 2016 IEEE Scott Helt Memorial Award, the 2016 IET Premium Award, the 2016 National Award for Science and Technology Progress (First Prize), the ICC2017 Best Paper Award, the 2018 IEEE ComSoc Asia-Pacific Outstanding Paper Award, and the 2020 IEEE ComSoc Leonard G. Abraham Prize. He was an Associate Editor of IEEE \textsc{Transactions on Wireless Communications} from 2011 to 2015 and IEEE \textsc{Communications Letters} from 2013 to 2016. He is currently an Associate Editor of IEEE \textsc{Transactions on Communications}, IEEE \textsc{Systems Journal}, and IEEE \textsc{Open Journal of Vehicular Technology}.
	
\end{IEEEbiography}

\begin{IEEEbiography}[{\includegraphics[width=1in,height=1.25in,clip,keepaspectratio]{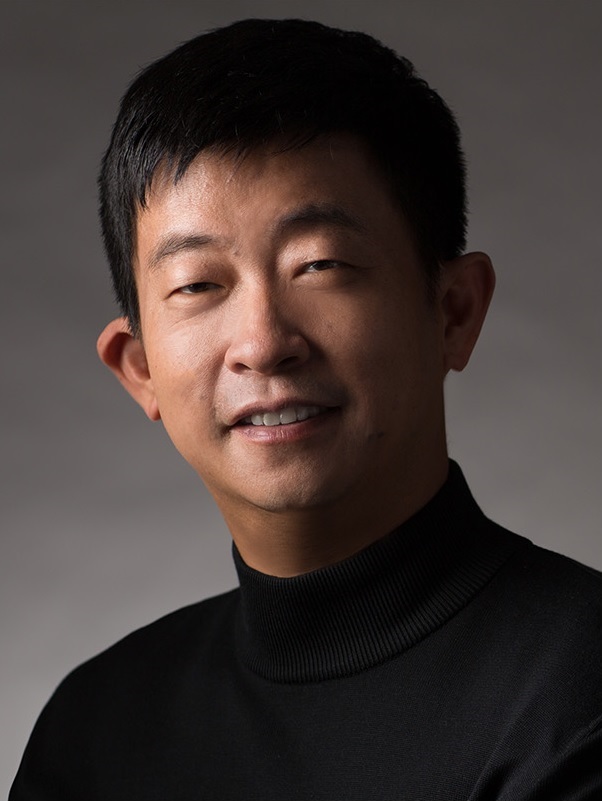}}]{Zhu Han}
(S’01–M’04-SM’09-F’14) received the B.S. degree in electronic engineering from Tsinghua University, in 1997, and the M.S. and Ph.D. degrees in electrical and computer engineering from the University of Maryland, College Park, in 1999 and 2003, respectively. 
	
From 2000 to 2002, he was an R\&D Engineer of JDSU, Germantown, Maryland. From 2003 to 2006, he was a Research Associate at the University of Maryland. From 2006 to 2008, he was an assistant professor at Boise State University, Idaho. Currently, he is a John and Rebecca Moores Professor in the Electrical and Computer Engineering Department as well as in the Computer Science Department at the University of Houston, Texas. Dr. Han’s main research targets on the novel game-theory related concepts critical to enabling efficient and distributive use of wireless networks with limited resources. His other research interests include wireless resource allocation and management, wireless communications and networking, quantum computing, data science, smart grid, carbon neutralization, security and privacy.  Dr. Han received an NSF Career Award in 2010, the Fred W. Ellersick Prize of the IEEE Communication Society in 2011, the EURASIP Best Paper Award for the Journal on Advances in Signal Processing in 2015, IEEE Leonard G. Abraham Prize in the field of Communications Systems (best paper award in IEEE JSAC) in 2016, IEEE Vehicular Technology Society 2022 Best Land Transportation Paper Award, and several best paper awards in IEEE conferences. Dr. Han was an IEEE Communications Society Distinguished Lecturer from 2015 to 2018 and ACM Distinguished Speaker from 2022 to 2025, AAAS fellow since 2019, and ACM distinguished Member since 2019. Dr. Han is a 1\% highly cited researcher since 2017 according to Web of Science. Dr. Han is also the winner of the 2021 IEEE Kiyo Tomiyasu Award (an IEEE Field Award), for outstanding early to mid-career contributions to technologies holding the promise of innovative applications, with the following citation: ``for contributions to game theory and distributed management of autonomous communication networks."
\end{IEEEbiography}

\end{document}